\title{Integral Action in Output Feedback \\
for multi-input multi-output nonlinear systems}

\author{Daniele Astolfi
 and Laurent Praly 
 \thanks{D. Astolfi is with CASY-DEI, University of Bologna, Bologna 40123,
Italy and with the MINES ParisTech, PSL Research University, CAS - Centre automatique et syst\'emes, Paris 75006, France, (e-mail: daniele.astolfi@unibo.it).}
\thanks{L. Praly is with the MINES ParisTech, PSL Research University, CAS - Centre automatique et syst\'emes, Paris 75006, France,
(e-mail: Laurent.Praly@mines-paristech.fr).}}

\documentclass[10pt,final,twocolumn]{IEEEtran}
\usepackage[dvipsnames]{xcolor}
\usepackage{graphics,epsfig,theorem,latexsym,amssymb,amsmath,picinpar,psfrag,amsfonts}
\usepackage{mathrsfs,bbding}
\usepackage{caption}
\usepackage{ulem}

\usepackage{ifthen}
\newboolean{Journal}
\setboolean{Journal}{false}
\newcommand{\IfJournal}[2]{\ifthenelse{\boolean{Journal}}{{#1}}{{#2}}}

\catcode`\@=11
\def\IEEEproofof#1{\par\noindent\hspace{2em}{\itshape Proof of #1: }}
\def\endIEEEproof{\hspace*{\fill}~\IEEEQED\par}
\catcode`\@=12

\newcommand{\hax}{{\hat{x}}}
\newcommand{\haX}{{\hat{X}}}

\def\bpsi{{\psi _{\hskip -0.7pt s\hskip -0.7pt a \hskip -0.7pt t}}}

\newcommand{\hphi}{{\hat{\phi }}}
\newcommand{\tphi}{{\tilde{\phi}}}

\newcommand{\alphalow}{\underline{\alpha}}
\newcommand{\alphaup}{\overline{\alpha}}
\newcommand{\vv}{\mbox{v}}
\def\vvs{{\mathchoice%
{{\mbox{ v}}}%
{{\mbox{\footnotesize v}}}%
{{\mbox{\scriptsize v}}}%
{{\mbox{\tiny v}}}%
}}

\newcommand{\ellmod}{\kappa}

\def\dotoverparen#1{\mathop{\vbox{\ialign{##\crcr\hfill$\cdot $\hfill\crcr
\noalign{\kern-2.3ex}
\downparenfill\crcr\noalign{\kern0.4ex\nointerlineskip}
$\hfil\displaystyle{#1}\hfil$\crcr}}}\limits}
\catcode`\@=12
\def\der{\mathaccent"7017 }

\newcommand{\RR}{\mathbb R}

\def\Stab{\mathcal{S}}
\def\Obs{\mathcal{O}}

\newcommand{\A}{\mathcal{A}}

\newcommand{\U}{\mathcal{U}}

\def\intGamma{%
\setbox0=\hbox{$\Gamma $}
\ht0 1.5ex
\vrule height 1.3em depth 0pt width 0pt
{\der{\box0}}
}

\def\intS{\der{\vrule height 0.7em depth 0pt width 0pt S}}

\def\intChatx{%
\der{%
\vrule height 0.7em depth 0pt width 0pt %
\mathcal{C} %
\vrule height 0pt depth 0pt width 0.2em%
}\hskip -0.2em{}_{\hax}%
}

\def\StabObs{\mathcal{S\!O}}

\def\hx{{\mathchoice%
{{\mbox{$\scriptstyle \mathcal{X}$}}}%
{{\mbox{$\scriptstyle \mathcal{X}$}}}%
{{\mbox{$\scriptscriptstyle \mathcal{X}$}}}%
{{\mbox{$\scriptscriptstyle \mathcal{X}$}}}%
}}

\catcode`\@=11
\def\downparenfill{$\m@th\braceld\leaders\vrule\hfill\bracerd$}
\def\overparen#1{\mathop{\vbox{\ialign{##\crcr\crcr \noalign{\kern0.4ex}
\downparenfill\crcr\noalign{\kern0.4ex\nointerlineskip}
$\hfil\displaystyle{#1}\hfil$\crcr}}}\limits}
\catcode`\@=12

\newcommand{\norm}[1]{\left| #1 \right|}

\newcommand{\pderiv}[2]{\dfrac{\partial #1}{\partial #2}}

\def\be{\begin{equation}}
\def\ee{\end{equation}}
\def\ba{\begin{array}}
\def\ea{\end{array}}
\def\bea{\begin{eqnarray}}
\def\eea{\end{eqnarray}}
\def\beann{\begin{eqnarray*}}
\def\eeann{\end{eqnarray*}}
\def\bmx{\begin{matrix}}
\def\emx{\end{matrix}}
\def\bpmx{\begin{pmatrix}}
\def\epmx{\end{pmatrix}}

\newtheorem{proposition}{Proposition}

\newtheorem{assumption}{Assumption}
\newtheorem{lemma}{Lemma}
\newtheorem{remark}{Remark}

\def\startmod{\color{black}}
\def\stopmod{\protect\normalcolor}

\newcounter{introduction}

\catcode`\@=11
\def\contenulabel{\protected@edef\@currentlabel}
\catcode`\@=12

\DeclareMathAlphabet\EuScript{U}{eus}{m}{n}
\SetMathAlphabet\EuScript{bold}{U}{eus}{b}{n}
\def\Comp{\EuScript{C}} 
\def\Lyap{\EuScript{V}} 

\def\s{\varsigma}

\begin{document}

\maketitle

\begin{abstract}
We address a particular problem of
output regulation  
for multi-input multi-output nonlinear systems. 
Specifically, we are interested in making the stability of an equilibrium point and 
the regulation to zero of an output, robust to
 (small) unmodelled discrepancies between design model and actual system
in particular those 
introducing an offset.
We propose
a novel procedure which
is intended to be
relevant to real life systems, as illustrated by a (non
academic) example. 
\end{abstract}

\begin{IEEEkeywords}
Robust regulation, nonlinear control, output feedback, semi-global stabilization,
integral action, observability, high-gain observer, forwarding, 
non-minimum phase systems,  uncertain dynamic system.
\end{IEEEkeywords}

\section{Introduction}

For a controlled dynamical system, it is of prime importance in 
real world applications to be able to design an output feedback control law which 
achieves asymptotic regulation of a given output while keeping the 
solutions in some prescribed set, in presence of
(constant) uncertainties. We refer to this as the problem of 
robust output regulation by output feedback.

The problem has been completely solved in the linear framework by Francis and Wonham
in the 70's (see \cite{Wonham}). Important efforts have been done in order to extend
this result to the nonlinear case (see, for instance 
\cite{Byrnes-Proscoli-Isidori,Isidori03}) and many different 
solutions have been proposed
 (see among others 
\cite{Chakrabortty07,Freeman96,Jiang01,Praly98},
\cite[Chapter~7.2]{Astolfi08}, \cite{anan85,huang90,Khalil00,Isidori03,Seshagiri-Khalil05, LiKhalil12}). 
Nevertheless we are still far from having a complete solution to the problem of output regulation
in the nonlinear multi-input-multi-output framework similar to what we have in the linear case. 
Indeed
most of the works require  
a good knowledge of the effects of the disturbances on the system, or
they \startmod rely \stopmod on ``structural properties'' as, 
for example, \textit{normal forms}, \textit{minimum phase assumption}, \textit{matched uncertainties} or \textit{relative degree
uniform in the disturbances}.
In particular, for single-input single-output minimum-phase nonlinear systems which possess a well defined relative degree 
preserved under the effect of disturbances,
a complete solution has been given in \cite{Khalil00}, further improved to the output feedback case 
in \cite{Seshagiri-Khalil01}. 
Under the same assumptions, this work has been successfully extended in \cite{Seshagiri-Khalil05}
to square multi-input multi-output systems for which
the notion of relative degree indices and observability indices coincides. 
Further,  with the technique of the 
\textit{auxiliary system} introduced in \cite{Isidori00}, the minimum-phase assumption has been removed in \cite{LiKhalil12}
allowing the \textit{zero-dynamics} to be unstable.
However, as far as we know, a general solution is still unknown when
these structural properties do not hold.

The approach to 
nonlinear output regulation followed in this paper is motivated
by the linear context developed in its full generality in the
milestone paper \cite{Francis76} that we find useful to briefly recall here.
Consider the linear system \startmod
$$
\ba{rcl}
\dot x &= &A_0x +B_0 u\,,
\\
y & = & C_0x \,,
\ea\quad 
y=
\left(\begin{array}{@{}c@{}}
y_r
\\
y_e
\end{array}\right)
=
\left(\begin{array}{@{}c@{}}
C_{0,r} \\ C_{0,e}
\end{array}\right) x\,,
$$
\stopmod
where the state $x$ is in $\RR^n$, the control $u$ is in $\RR^m$ and
the measured output $y$ is in $\RR^p$. The output
$y$ is decomposed as $y=(y_r,y_e)$
where $y_r$, in $\RR^r$,  $r\leq m$, is the output to be regulated to zero (without loss of generality). 
When the system above is 
supposed to be only an approximation of a process given by
\startmod
$$\!\!
\ba{rcl}
\dot x \!&= &\!A x +B u + P w\,,
\\
y \!&= &\!Cx +Qw\,,
\ea \;\,
y=\left(\begin{array}{@{}c@{}}
y_r
\\
y_e
\end{array}\right)
=
\left(\begin{array}{@{}c@{}}
{C}_r \\ {C}_e
\end{array}\right) x+ \left(\begin{array}{@{}c@{}}
{Q}_r \\ {Q}_e
\end{array}\right) w\,,
$$
\stopmod
where $w$ \startmod is an unknown constant signal to be either rejected or tracked\stopmod, the \textit{well posed regulator problem with internal 
stability} (addressed by Wonham for linear systems as shown for instance in
\cite[Chapter 8]{Wonham}) is that of finding an output feedback law 
based on the model 
such that, for all triplets \startmod $\left\{{A},{B},C \right\}$ close enough to
$\left\{A_0,B_0,C_0 \right\}$, and for all matrices pairs 
$\left\{{P},Q\right\}$,  \stopmod
the regulation-stabilization problem is solved, i.e. the system admits a stable equilibrium point
on which the output to be regulated is equal to zero. 
According to \cite[Proposition 1.6]{Byrnes-Proscoli-Isidori}, for example,
this problem is solvable if and only if the following $3$ conditions are 
satisfied:
\begin{list}{}{%
\parskip 0pt plus 0pt minus 0pt%
\topsep 0.5ex plus 0pt minus 0pt%
\parsep 0pt plus 0pt minus 0pt%
\partopsep 0pt plus 0pt minus 0pt%
\itemsep 0.5ex plus 0pt minus 0pt
\settowidth{\labelwidth}{(a)}%
\setlength{\labelsep}{0.5em}%
\setlength{\leftmargin}{\labelwidth}%
\addtolength{\leftmargin}{\labelsep}%
}\startmod
\item[(a)] 
the pair $\left(A_0,C_0\right)$ is detectable;
\item[(b)]
the pair $(A_0,B_0)$ is stabilizable;
\item[(c)]
the matrix $\left(\begin{array}{@{}cc@{}}
A_0 & B_0
\\
C_{0,r} & 0
\end{array}\right)$ is right invertible. \stopmod
\end{list}
Precisely, under the above $3$ conditions, it is always possible to design an output feedback law of the form
\[\ba{rcl}
\dot z & = & y_r \\
\dot \eta & = & F\eta + L y \\
u & = & K \eta + M z + N y 
\ea\]
which solves the regulation problem provided $F$, $L$, $K$, $M$, and 
$N$ are chosen such that the following matrix  \startmod
\[
\bpmx A + B N C & B K & B M
\\
L C  & F & 0 
\\
C_r & 0 & 0 \epmx
\] 
 is Hurwitz
for all triplets  $\left\{{A},{B},C \right\}$ close enough to
$\left\{A_0,B_0,C_0 \right\}$, and for all matrices pairs 
$\left\{{P},Q\right\}$, \stopmod
Note that \startmod in this linear framework \stopmod no \textit{structural properties} are needed.

Merging all the tools in literature that are at our disposal, we try to recover the same result
as in the linear case,
asking for possibly minimal assumptions but at the same time paying particular attention to proposing a 
design truly workable in applications. For example, 
minimality implies not to ask for any specific structural 
properties whereas applicability forbids
nonlinear changes of coordinates when no expression is known for their inverse. 
Our answer to the problem uses  ``bricks''
which can be found in other publications (as \cite{TeelPraly94,PoulainPraly10,AstolfiPraly12})
that we glue together. But for making this glueing process efficient we have to address some (new) specific
problems.

As in the linear framework, we extend the system with an integral action.
Then, as in \cite{PoulainPraly10}, we rely on forwarding to design a stabilizing state feedback
for the extended system. Next, for transforming this state feedback 
into an ouput feedback, it is sufficient to apply the techniques which have been proposed for 
asymptotic stabilization by output feedback. A lot of effort has 
been devoted to this question and many results have accumulated (see for instance 
the survey \cite{Andrieu-Praly09}).
In particular the transformation is done by replacing the actual state by a state estimate 
provided by a tunable observer (i.e. an observer whose dynamics can be made
arbitrarily fast). Stability of the overall closed loop system is established via the common
separation principle \cite{TeelPraly94,Atassi-Khalil},
and output regulation follows from the integral
action embedded in the control law.

The tunable observer we propose is, as in  \cite{AstolfiPraly12} (previously inspired by \cite{Deza} and
\cite{Maggiore-Passino}), a high-gain observer written in the original
coordinates and appropriate for our multi-input multi-output (possibly non-square) case.
We propose a new set of sufficient conditions which guarantees 
the existence of such an observer. As opposed to what we have found in the literature
(see for instance  \cite{Bornard,HammouriBornard10, GK}) our conditions 
can be verified in the original coordinates and they do not need the explicit knowledge
of the inverse of nonlinear change of coordinates (which may be very hard to find).
Also in looking for minimal assumptions, we do not ask for global observability or global uniformity with respect to the 
inputs. 
The latter impacts the state feedback design and we show how to address this point
(in \cite{PoulainPraly10} only a global solution is proposed). 

Finally, we show that the proposed solution guarantees robust regulation. 
Robustness is here with respect to unmodelled effects, not in the system state
dimension, but in the approximations of the functions which define its
dynamics and measurements. 
This has been done already in \cite{PoulainPraly10} but for the state feedback case and with an assumption on the closed 
loop system.
Here we show that 
if the model is close enough (in a $C^1$ sense), in open loop,
to the process, then output regulation is achieved by our output feedback design. 
However, as opposed to the linear
case, where the result is global with respect to the magnitude of the 
disturbances, an unfortunate consequence of being in
our less restrictive context is that we need the perturbations to be 
small enough. 

In this work, to simplify, we restrict our attention to
systems affine in the input. The extension to the non affine case is
made possible by considering the system controls as state 
and their derivatives as fictitious controls. See \cite{AstolfiPraly12} 
for example.

The paper is organized as follows.
Section \ref{sec:mainresult} is devoted to show the main assumption and results 
of this work.
In Section \ref{sec:statefeedback} and \ref{sec:observer} we present respectively the state feedback design 
and the observer design.
The proofs of the main propositions are given in Section \ref{sec:proofs}.
Finally, in Section \ref{sec:example}, we illustrate the proposed design with a non-academic example 
inspired
from a concrete case study in aeronautics (the regulation
of the flight path angle of a simplified longitudinal model of
a plane). 

%
%

\subsection*{Notations}
For a set $S$, $\intS$ denotes its interior, $\partial S$ denotes its boundary
and $d(x,S)$ denote the distance function of a point $x$ to the set $S$. 
When $S$ is a subset of $\mathcal{A}\times \mathcal{B}$ whose points are 
denoted $(a,b)$, 
$(S)_a$
denotes the set $\{a\in \mathcal{A}:\,  \exists 
b\in \mathcal{B}: (a,b)\in S\}$.
For a function $h$ and a vector field $f$, $L_f$ denotes the Lie 
derivative of $h$ along $f$, given coordinates $x$, $L_f h(x) = \pderiv hx (x) f(x)$.
To any strictly positive real number $v$, we associate a 
``saturation''  function 
$\texttt{sat}_{v} $ defined as a $C^1$ function bounded by $v$ 
and satisfying
\be\label{eq:sat}
\texttt{sat}_{v} (s)
 = s \qquad \mbox{if}\quad  |s|\leq \dfrac{v}{1+\s}\  ,
\ee
where $\s$ is a (small) strictly positive real number.

\section{Robust Regulation by Output Feedback}
\label{sec:mainresult}

\subsection{Problem Statement and Assumptions}
\label{sec:statement}
For a process, we have at our disposal the following dynamical model
\begin{align}
\label{eq:system}
\dot x &= f(x) + g(x)u, & y &= h(x)=(h_r(x),h_e(x))
\end{align}
where the state $x$ is in $\RR^n$, the control $u$ is in $\RR^m$,
the measured output $y$ is in $\RR^p$  and the functions
$f: \RR^n \to \RR^n$, $g: \RR^n \to \RR^{nm}$ and $h: \RR^n \to 
\RR^p$ are smooth enough and \startmod $f$ and $h$ are \stopmod zero at the origin.
We investigate the problem of regulating at zero the part $y_r$ 
of the output $y$ decomposed as $y=(y_r,y_e)$
with $y_r\in\RR^r$ and $r\leq m$ and this while stabilizing an 
equilibrium for $x$. But, being aware that the triplet $(f,g,h)$ 
gives only an approximation of the dynamics of the process, we would 
like the above regulation-stabilization property to hold not only 
for this particular triplet but also for any other one in a 
neighborhood.

The real process  is described by equations 
of the form
\begin{align}
\label{eq:process}
\dot x &= \xi(x,u), & y &= \zeta(x,u),
\end{align}
where the functions $\xi:
\RR^n \times \RR^m \to \RR^n$ and $\zeta: \RR^n \times \RR^m \to
\RR^p$ are assumed continuously differentiable $(C^1)$. These 
functions are unknown but we assume that they are close enough to
$f+gu$ and $h$ respectively in the sense that the discrepancies
$$
\left|\xi(x,u) - f(x)-g(x)u\right|
\;+\; 
\left|\zeta(x,u)-h(x)\right|
$$
and
$$
\left|\left(\begin{array}{cc}
\frac{\partial \xi}{\partial x}(x,u)-\frac{\partial f}{\partial 
x}(x) -\frac{\partial g}{\partial x}(x)u
&
\frac{\partial \xi}{\partial u}(x,u)-g(x)
\\
\frac{\partial \zeta}{\partial x}(x,u)- \frac{\partial h}{\partial x}(x)
&
\frac{\partial \zeta}{\partial u}(x,u)
\end{array}\right)\right|
$$
are small enough as made precise later on.

Mimicking the $3$ necessary and sufficient conditions for the linear 
case given in the introduction,
we consider the following (sufficient) assumptions that we discuss 
after their formal statement.

\begin{assumption}\label{ass:observer}
There exists
an open set $\Obs$ of $\RR^n$ containing the origin and an open star-shaped subset
$\U$ of $\RR^m$, with the origin as star-center,
such that, for any strictly positive real number $\bar u$
and any
compact subset $\mathfrak{C}$ of $\Obs $,
there exist
an integer $d$,
a compact subset $\widehat{\mathfrak{C}}$ of $\Obs $, a real
number $\overline{U}$ and
a class-$\mathcal{K}^\infty$ function $\alphalow $ such that, for each
each integer $\ellmod$,
we can find $C^1$ functions $\vartheta_\ellmod:\RR^m\times\RR^p\times\Obs  \to \Obs $,
$U_\ellmod:
\Obs  \times \Obs 
\to\RR_{\geq 0}$, a continuous function $L_{\vartheta_\ellmod}:\Obs\to \RR_{\geq 0}$ and a strictly positive real number
$\sigma_\ellmod$, such that:
\begin{list}{}{%
\parskip 0pt plus 0pt minus 0pt%
\topsep 0.5ex plus 0pt minus 0pt%
\parsep 0pt plus 0pt minus 0pt%
\partopsep 0pt plus 0pt minus 0pt%
\itemsep 1ex plus 0pt minus 0pt
\settowidth{\labelwidth}{1.}%
\setlength{\labelsep}{0.5em}%
\setlength{\leftmargin}{\labelwidth}%
\addtolength{\leftmargin}{\labelsep}%
}
\item[1.]
for any function $t\to u(t)$ with values in $\U(\bar u)$ defined as
\be\label{eq:Uu}
\U(\bar u) = \left\{u\in \U\,  :\, |u|\leq \bar u \right\}
\ee
and any bounded function $t\to y(t)$,
the set $\widehat{\mathfrak{C}}$ is forward invariant by the
flow generated by the following observer
\begin{equation}
\label{eq:observer}
\dot \hax_\ellmod = \vartheta_\ellmod (y,\hax_\ellmod,u)
\;;
\end{equation} 
\item[2.]
  $\displaystyle
\forall (x,\hax)\in\Obs \times\Obs \,, \quad U_\ellmod (x,\hax)=0
\quad \Longleftrightarrow \quad x=\hax
\;;
$\hfill \null
\item[3.]
  $\displaystyle
\sigma_\ellmod^{-d}\,  \alphalow(|x-\hax|)
\; \leq  \; U_\ellmod\,(x,\hax)
\; \leq \; \sigma_\ellmod^{d}\, \overline{U} $\hfill \null \refstepcounter{equation}\label{eq:assobs2}$(\theequation)$
\begin{flushright}
$ \forall x\in \mathfrak{C}
\: ,\; \forall \hax  \in \widehat{\mathfrak{C}}\;;\quad
$
\end{flushright}
\item[4.]   $\displaystyle
\lim_{\ellmod\to\infty} \sigma_\ellmod = +\infty
\  ;
$\hfill \null 
\refstepcounter{equation}\label{eq:LP29}$(\theequation)$
\item[5.]
$\displaystyle
\frac{\partial U  _\ellmod}{\partial x}
(x,\hax_\ellmod) 
\!
\left[f(x)\!+\!g(x)u\right]
$\hfill \null 
\refstepcounter{equation}\label{eq:assobs1}$(\theequation)$
\\\null \hfill $\displaystyle
+ \; \frac{\partial U  _\ellmod}{\partial \hax_\ellmod}
(x,\hax_\ellmod) \vartheta_\ellmod (h(x),\hax_\ellmod,u)\: \leq \: -\sigma_\ellmod\,U_\ellmod\,(x,\hax_\ellmod)
$\hfill \null \\\null \hfill 
$\displaystyle \qquad\qquad\quad
\forall \, u\in \U(\bar u),\quad
\forall \,(x,\hax_\ellmod)\in
\mathfrak{C}\times \widehat{\mathfrak{C}}
\  .$
\item[6.]
For all $(y_a,y_b,\hax_\ellmod,u)$ in
$\RR^{2p}\times \Obs\times \U(\bar u)$,
\be\label{eq:LP23}
\left|\vartheta_\ellmod (y_a,\hax _\ellmod,u)
-
\vartheta_\ellmod (y_b,\hax _\ellmod,u)\right|
\; \leq \; L_{\vartheta_\ellmod} (\hax _\ellmod)\,  |y_a-y_b|
\ee

\end{list}
\end{assumption}

\begin{assumption}
\label{ass:stability}
There exist an open subset $\Stab$ of $\RR^n$ and a continuous function 
$\beta: \Stab \to \U$ which is 
zero at the origin and such that the origin of \eqref{eq:system}
with $u=\beta (x)$,
is an asymptotically and locally exponentially stable equilibrium point with $\Stab$ as domain of attraction.
\end{assumption}

\stopmod

\begin{assumption}
\label{ass:staticgain}
The matrix
\begin{equation}
\left(\begin{array}{@{}cc@{}}
\frac{\partial f}{\partial x}(0) & g(0)
\\[0.5em]
\frac{\partial h_r}{\partial x}(0)  & 0
\end{array} \right)
\end{equation}
is right invertible.
\end{assumption}

Assumption \ref{ass:observer} is aimed at being a counter-part
of the detectability condition (a). But we have to face here problems 
specific \startmod to this nonlinear framework: \stopmod
\begin{itemize}{}{%
\parskip 0pt plus 0pt minus 0pt%
\topsep 0.5ex plus 0pt minus 0pt%
\parsep 0pt plus 0pt minus 0pt%
\partopsep 0pt plus 0pt minus 0pt%
\itemsep 0.5ex plus 0pt minus 0pt
\settowidth{\labelwidth}{iii)}%
\setlength{\labelsep}{0.5em}%
\setlength{\leftmargin}{\labelwidth}%
\addtolength{\leftmargin}{\labelsep}%
}
\item In our construction we shall rely on
the so called separation principle. For nonlinear systems (see \cite{TeelPraly94} 
for example), it asks for an observer with a tunability property, i.e. an observer the
speed of convergence of which can be made arbitrary fast (see 
\cite{Besancon07}). This property is provided here by the family of
observers (\ref{eq:observer}) satisfying
(\ref{eq:assobs2}), (\ref{eq:assobs1}) and (\ref{eq:LP29}).
\item
Observability may depend on the input. This explains why we impose the 
control to \startmod belong \stopmod to the set $\U$.
\item
The tuning of observers for non linear systems may depend on the 
local Lipschitz constant of the non linearities. This explains why 
the family of observers depends on the bound $\bar u$ of the input.
\end{itemize}
On the other hand, to reduce the restrictiveness, Assumption 
\ref{ass:observer} is imposed only for system states belonging to an open subset $\Obs $ of 
$\RR^n $. In Section \ref{sec:observer} we shall see how the family 
of observers in this assumption can be designed as observers based
on high-gain techniques.

Assumption \ref{ass:stability} is the counter-part of the 
stabilizability condition (b) and claims the existence of a  
state feedback law
which asymptotically stabilizes the system \eqref{eq:system}. Actually 
it assumes \startmod that a preliminary design step can be done. \stopmod For it any tool 
-- Lyapunov design, feedback (partial) linearization, passivity, use of 
structure of uncertainties in combination with gain assignment 
techniques, etc. -- can be exploited. 
However, because Assumption \ref{ass:observer} imposes the 
control be in $\U$, we propagate this restriction here, asking the 
stabilizing control $\beta $ to take values in that set.
On the other hand, 
we can cope with having an arbitrary domain of 
attraction $\Stab$, no need for it to be the full space or any 
arbitrarily large compact set.

Finally Assumption \ref{ass:staticgain} corresponds to the non-resonance condition (c)
and states that the first order approximation at the origin
of the system \eqref{eq:system} does not have any zero at $0$.

\subsection{Main results}

Assumptions 1 to 3 are sufficient to guarantee the existence of an 
output feedback law solving the regulation-stabilization problem for the 
model.

\begin{proposition}\label{prop:outF}
Suppose Assumptions  \ref{ass:observer}, \ref{ass:stability} and \ref{ass:staticgain} 
 hold. There exists
an open subset $\StabObs  $ 
of $(\Stab\cap \Obs )\times \RR^r$ such that, for
any of its compact set $\mathcal{C}_{x\!z}$, there
exist
an integer $\underline{\ellmod}$, a compact subset
$\mathcal{C}_{\hax}$ of $\Obs $,
a real number $\mu$ 
and $C^1$ functions
$k:\RR^n\times\RR^r\to\RR^r$  and
$\bpsi:
\RR^n\times\RR^r
\to\U(\mu)$, such that the origin of the model
\eqref{eq:system}, in closed-loop with the dynamic output 
feedback
\be\label{eq:controller}
\dot z = k(\hax, h_r(x))\;, \quad
\dot \hax = \vartheta_\ellmod (y,\hax,u)
\; ,\quad
u= \bpsi(\hax,z) 
\ee
with $\ellmod\geq \underline{\ellmod}$,
is asymptotically stable with a
domain of attraction
$\A$ containing the set $\intChatx\times \mathcal{C} _{x\!z}$.
\end{proposition}

\IfJournal%
{\begin{IEEEproof} \startmod
This proof 
follows the same 
lines as in \cite{AstolfiPraly12}, inspired by \cite[Chapter 
12.3]{Isidori2}. We omit it to save 
space. It can be found in \cite{AstolfiPraly15}. \stopmod
\end{IEEEproof}}
{\begin{IEEEproof}
See Section \ref{sec:proofOutF}.
\end{IEEEproof}
}

\par\vspace{1em}
In the case where $\Stab$ and $\Obs $ are the full space $\RR^n$, this result would 
be a semi-global regulation-stability result. It claims the existence 
of a dynamic output feedback which 
asymptotically stabilizes the origin 
of the model \eqref{eq:system}. Such a result is not new per se. It is in line with many results 
related to the separation principle as those in
\cite{TeelPraly94}, \cite{Atassi-Khalil} or \cite[Chapter 
12.3]{Isidori2}.

But as written in the introduction,
we do \startmod
not state only ``existence'' but instead we propose an explicit and workable 
design.   \stopmod
\IfJournal%
{For example, we}
{We}
refer the reader to Section 
 \ref{sec:statefeedback} for
the definition\footnote{%
See respectively (\ref{eq:setSO}) and (\ref{eq:LP13}) for $\StabObs$, \eqref{eq:LP15} for $\mu$, 
(\ref{eq:LP3}) and \eqref{eq:integr} for $k$
and \eqref{eq:bpsi} for $\bpsi$. }
of the set $\StabObs$,  the real number $\mu$ and the functions $k$  
\IfJournal{}{
  and
$\bpsi$ and to Section \ref{sec:proofOutF} for the 
definition\footnote{%
See successively (\ref{eq:LP16}), (\ref{eq:defChatx}).}
of the integer  $\underline{\ellmod}$ and the set 
$\mathcal{C}_{\hax}$
}.

\startmod In the following propositions, under the Assumptions  \ref{ass:observer}, \ref{ass:stability} and \ref{ass:staticgain} 
and knowing the result of Proposition \ref{prop:outF} holds \stopmod, we study the process (\ref{eq:process})
in closed-loop with the control law  \eqref{eq:controller} designed for the model (\ref{eq:system}).


\begin{proposition}\label{prop:robust}
Let $\overline{\Comp}$ be an arbitrary compact subset  of the domain of attraction $\A$, 
given by Proposition \ref{prop:outF}, which admits the 
equilibrium as an interior point and is forward 
invariant for the closed-loop system 
(\ref{eq:system}),(\ref{eq:controller}). For any open neighborhood 
$\mathcal{N}_{\partial \overline{\Comp}}$ of the boundary set
$\partial \overline{\Comp}$, contained in $\A$, there exists a strictly 
positive real number $\delta $ such that, for any pair $(\xi ,\zeta 
)$ of $C^1$ functions which satisfies
\begin{multline} 
\label{eq:LP2}
\left|
\vrule height 0.5em depth 0.5em width 0pt
\xi(x, u) -[f(x) + g(x)u]
\right| 
+ 
\left|
\vrule height 0.5em depth 0.5em width 0pt
\zeta (x,u)-h(x)\right| \leq  \delta 
\\
\forall \,(x,u)\in \,\left(\mathcal{N}_{\partial \overline{\Comp}}\right)_x \times \U(\mu)
\end{multline}
the closed-loop system \eqref{eq:process}, \eqref{eq:controller}
has equilibria and at any such point the output $y_r$ is zero.
\end{proposition}

\begin{IEEEproof}
See Section \ref{sec:proofRobust}.
\end{IEEEproof}
\par\vspace{1em}

If the domain of attraction were the full space, this result would 
follow from \cite[Section 12]{Sontag}. It says that, when the 
evaluation, on a 
``spherical shell''-like set, of the model and 
process functions are close enough, equilibria where 
output regulation occurs do exist. If this closeness is everywhere in 
the domain of attraction, then we have even a solution to the the well posed regulator problem with internal 
stability.

\begin{proposition}\label{prop:robust2}
For any compact sets $\underline{\Comp}$ and $\overline{\Comp} $, the latter 
being forward invariant 
for the closed-loop system (\ref{eq:system}),(\ref{eq:controller}), which satisfy
$$
\{0\}\subsetneqq \underline{\Comp}\subsetneqq \overline{\Comp} \subsetneqq  \A
\  ,
$$
and for any open neighborhood 
$\mathcal{N}_{ \overline{\Comp}}$ of 
$\overline{\Comp}$, contained in $\A$, 
there exists
a strictly positive real number $\delta$ such that, 
to any pair $(\xi ,\zeta )$ of $C^1$ functions which satisfies
\begin{multline} 
\label{OF1}
\left|
\vrule height 0.5em depth 0.5em width 0pt
\xi(x, u) -[f(x) + g(x)u]
\right| 
+ 
\left|
\vrule height 0.5em depth 0.5em width 0pt
\zeta (x,u)-h(x)\right| \leq  \delta 
\\
\forall \,(x,u)\in \, \startmod \overline{\Comp}_x \stopmod \times \U(\mu)
\end{multline}
and 
\\[0.7em]$\displaystyle 
\left| \!
\left(\begin{array}{@{}c@{\quad }c@{}} 
\pderiv \xi x  (x,u)&
\pderiv \xi u (x,u)\\[1em]
\pderiv \zeta x (x,u)
& 
\pderiv \zeta u (x,u)
\end{array}\right)
- \left(\begin{array}{@{}c@{\quad }c@{}}
\pderiv fx (x)+ \pderiv gx (x) u & g (x)\\[1em]
 \pderiv hx  (x)& 0
\end{array}\right)\!
 \right| 	
$\hfill \null \\[0.5em]\null\hfill$\displaystyle 
\le\;  \delta \qquad \forall (x,u) \in \underline{\Comp}_{x} \times \U(\mu) \  ,
$\refstepcounter{equation}\label{OF2}\qquad $(\theequation)$
\\[0.7em]
we can associate
a point
$\hx_e = (x_e, z_e, \hax_e)$ which is an exponentially stable 
equilibrium point of \eqref{eq:process}, \eqref{eq:controller}
whose basin of attraction $\cal B$ contains $\overline{\Comp}$.
Moreover, any solution $(\haX(\hx,t),X (\hx,t),Z(\hx,t))$ of  \eqref{eq:process}, \eqref{eq:controller}
with initial condition $\hx$ in ${\cal B} $ satisfies
\begin{equation}
\label{OF5}
\lim_{t\to +\infty} \zeta_r\left(
\vrule height 0.75em depth 0.75em width 0pt
X(\hx,t),\bpsi\left(
\haX(\hx,t),Z(\hx,t)
\right)\right) = 0\;.
\end{equation}
\end{proposition}

\begin{IEEEproof}
See Section \ref{sec:proofRobust2}.
\end{IEEEproof}
\par\vspace{1em}

This statement is of the same spirit as those claiming that under the 
action of (small) perturbations, asymptotic stability is transformed 
into semiglobal practical stability. But here we have more since we 
have existence of a single equilibrium for which the regulated output is zero.
And for this no specific structure of the 
unmodelled effects is required. 

\section{State Feedback Design}
\label{sec:statefeedback}

\subsection{Adding an integral action: design of the function $k$}
\label{sec:integral}
To solve the problem of regulating $y_r$ to 0
we follow the very classical idea of adding an integral action.
To do so we first select a $C^1$ function 
$k:\RR^n\times\RR^r\to\RR^r$ satisfying\footnote{%
When $L_gL_f^ih_r(x)=0$, for $i$ in 
$\{0,\ldots,\rho \}$, (\ref{eq:kcond1}) can be relaxed in
$\left\{
\vrule height 0.5em depth 0.5em width 0pt
k(x,h_r(x)) = 0 ,\,  L_fh_r(x)=\ldots=L_f ^{\rho -1}h_r(x)=0
\right\}\  \Rightarrow \   h_r(x) =0$.
See \cite{Seshagiri-Khalil05} for example.
}, for all $x$ in $\RR^n$ and all $(y_{r}^a,y_r^b)$ in $\RR^{2r}$,
\begin{eqnarray}
\label{eq:kcond1}
&\displaystyle
k(x,y_r) = 0 \qquad \Leftrightarrow\qquad y_r=0
\  ,
\\
\label{eq:kcond2}
&\displaystyle
|k(x,y_{r}^a)-k(x,y_{r}^a)|\; \leq \; L_k(x)\,  |y_{r}^a-y_{r}^b|
\  ,
\end{eqnarray}
where $L_k:\RR^n\to \RR_{\geq 0}$ is a continuous function.
Of course the function $k$ can
be simply $h_r$. But, in its choice, we can take advantage of the properties of the physical system under 
consideration and it can simplify the feedback design or its 
implementation. \startmod An example is given  \stopmod in section 
\ref{sec:example}. For the time being note that smoothness of $k$ and 
(\ref{eq:kcond1}) implies
\begin{equation}
\label{eq:LP21}
\frac{\partial k}{\partial x}(x,0)\;=\; 0
\qquad \quad \forall \, x\in \RR^n
\end{equation}
and  the existence\footnote{%
The function $\eta $ is a smoothened version of
$s\to\sup_{(x,y_r): |k(x,y_r)|\leq s}\frac{|y_r|}{1+|x|+|y_r|^2}$.}
of a continuous function 
$\eta :\RR_{\geq 0}\to \RR_{\geq 0}$ satisfying $\eta (0)=0$ and
\begin{equation}
\label{eq:LP22}
|y_r|\; \leq \; [1+|x|+|y_r|^2]\,  \eta (|k(x,y_r)|)
\qquad \forall (x,y_r)\in \RR^n\times \RR^p
\  .
\end{equation}
Actually the function $k$ used in the output feedback (\ref{eq:controller}) 
is the modified version given later in (\ref{eq:satbarx}).

\subsection{Design the function $\psi$ via forwarding}
\label{sec2}
Let us consider the extended system 
\be\label{eq:extended}
\dot x = f(x) + g(x)u, \qquad \dot z = k(x,h_r(x)) \  .
\ee
With Assumption \ref{ass:stability}, we are left with modifying the given 
state feedback $\beta $ to obtain a state feedback stabilizing 
asymptotically the origin for the extended system (\ref{eq:extended}). Fortunately it has
the so-called feedforward form which
has been extensively studied in the 90's
with in particular the introduction of the
 forwarding techniques based on saturations as in \cite{Teel96} or on Lyapunov
design with coordinate change as in \cite{Mazenc96} or coupling term 
as in
\cite{Jankovic96}.  We recall briefly these techniques. 
They differ on the available knowledge
they require.
Specifically, Assumption \ref{ass:stability} has two
consequences~:
\begin{list}{}{%
\parskip 0pt plus 0pt minus 0pt%
\topsep 0.5ex plus 0pt minus 0pt%
\parsep 0pt plus 0pt minus 0pt%
\partopsep 0pt plus 0pt minus 0pt%
\itemsep 1ex plus 0pt minus 0pt
\settowidth{\labelwidth}{1.}%
\setlength{\labelsep}{0.5em}%
\setlength{\leftmargin}{\labelwidth}%
\addtolength{\leftmargin}{\labelsep}%
}
\item[1.]
With the converse Lyapunov theorem of \cite{Kurzweil56}, we know
there exists a $C^1$ function $V:\Stab \to \RR_{\geq 0}$ which is positive definite and proper
on $\Stab$ and such that the function $x \mapsto \pderiv Vx
(x)
\Big( f(x) + g(x)\beta(x) \Big)$ is negative definite
on $\Stab$ and upperbounded by a negative definite quadratic form of $x$ in a 
neighborhood of the origin.
\item[2.]
Since the origin of the system \eqref{eq:system} in closed-loop with $\beta(x)$ is locally exponentially stable, there exists 
(see \cite[Lemma IV.2]{Mazenc96}) a $C^1$ function $H: \Stab \to \RR^r$ satisfying
\begin{equation}
\label{eq:beta}
\pderiv H x(x) \left(f(x) + g(x)\beta(x) \right) = k(x,h_r(x))\: ,\;  H(0) = 0.
\end{equation}
\end{list}
Depending on whether or not we know the function $V$ and/or the
function $H $ or only its first order approximation at the origin leads to different designs.
\begin{list}{}{%
\parskip 0pt plus 0pt minus 0pt%
\topsep 1ex plus 0pt minus 0pt%
\parsep 0pt plus 0pt minus 0pt%
\partopsep 0pt plus 0pt minus 0pt%
\itemsep 2ex plus 0pt minus 0pt
\settowidth{\labelwidth}{\  a)}%
\setlength{\labelsep}{0.5em}%
\setlength{\leftmargin}{\labelwidth}%
\addtolength{\leftmargin}{\labelsep}%
}
\item[\  a)]
\textit{Forwarding with $V$ and $H$ known}\\[0.3em]
When both $V$ and $H$ are known, a stabilizer $\psi $ for the system \eqref{eq:extended},
 is
\\[0.7em]$\displaystyle 
\psi (x,z)\;=\; 
$\refstepcounter{equation}\label{eq:forwarding}\hfill$(\theequation)$
\\[0.3em]\null\hfill$\displaystyle 
\beta(x) - J\left(x, \left[ L_gV (x) - (z - H(x))^\top L_gH (x) 
\right]^\top \right) ,
$\\[0.7em]
with $H$ defined by \eqref{eq:beta}, and with $J:\RR^n\times 
\RR^m\to\RR^m$ any continuous function satisfying,
for all $x\in\RR^n$,
\begin{equation}
\label{eq:Upsilon}
v^\top J(x,v) > 0 \quad \forall \,v \neq 0
\  ,\quad    \det\left( \pderiv J v (x,0)\right) \ne 0.
\end{equation}
Following \cite{Mazenc96} this can be established
under Assumptions
\ref{ass:stability} and \ref{ass:staticgain}
 with the function
$V_e:\Stab\times\RR^r \to \RR_{\geq 0}$ defined as
\begin{equation}
\label{eq:lyapunov1}
V_e(x,z) = V(x) + \frac12 (z - H(x))^\top (z - H(x)).
\end{equation}

\begin{remark}
\label{rem1}
If $V$ is known from the design of $\beta $, it may not be proper
on $\Stab$. To make it proper we first define $v_\Stab $ as
$$
v_\Stab=\inf_{x\not\in \Stab} V(x)
$$
and we replace $V (x)$ by $\frac{V (x)}{v_\Stab-V(x)}$. See [32]. Unfortunately in
doing so, the domain of definition of this new function $V$
may be a strict subset of $\Stab$. In the following, we still call $\Stab$ this 
domain on which $V$ is proper.
\end{remark}


\item[\  b)]
\textit{Forwarding with $V$ unknown but $H$ known}\\[0.3em]
\label{sec:forwarding_betav}
When $V$ is unknown, but $H$ is known, there exists a
 function $\gamma : \Stab \to
 \RR_{\geq 0}
 $ with strictly positive values
 such that a state feedback for the system \eqref{eq:extended} is
\begin{equation}
\label{eq:perturbationobserver}
\psi (x,z) = \beta(x) + \gamma (x) L_gH(x) ^\top J(x, z - H(x)),
\end{equation}
with $H$ defined by \eqref{eq:beta}, and 
$J:\RR^n\times\RR^r\to\RR^r$
bounded and satisfying \eqref{eq:Upsilon}.  This can
be established with the Lyapunov function \eqref{eq:lyapunov1}.


\item[\  c)]
\textit{Forwarding with $V$ unknown and $H$ approximated}\\[0.3em]
Instead of solving the partial differential equation \eqref{eq:beta}
for $H$, and using (\ref{eq:perturbationobserver}), we pick
\begin{equation}
\psi (x,z) = \beta(x) + \gamma (x) g(x)^\top {H_0}^\top J(x, z -
{H_0}x)\  ,
\end{equation}
where $H_0$ is obtained as
\begin{equation}
\label{eq:beta0}
H_0
= 
\frac{\partial k}{\partial y_r}(0,0)\frac{\partial h_r}{\partial x}(0)
\left[\frac{\partial f}{\partial x}(0)
+ g(0)\frac{\partial \beta }{\partial x}(0)
\right]^{-1}
\!\!\!\!.
\end{equation}

The corresponding Lyapunov function is
\begin{equation*}
V_e(x,z) = d(V(x)) + \sqrt{1 + \tfrac12 (z - H_0x)^\top (z - H_0x)} - 1,
\end{equation*}
where $d:\RR_{\geq 0} \to \RR_{\geq 0}$ is a $C^1$ function with
strictly positive derivative, to be chosen large enough (see
\cite{Mazenc96}).

In the case where the system
\begin{equation}
\dot x = f(x) + g(x)(\beta(x) + v),
\end{equation}
with $v$ as input is input to state stable with restriction, i.e. provided $\norm v$ is bounded by some given
strictly positive real number $\Delta$, then following \cite{Teel96}, the state feedback can be chosen as
\begin{equation}
 \psi (x,z)  = \beta(x) + \epsilon J\left(x, \frac{g(0)^\top H_0^\top (z - H_0x)}{\epsilon} \right),
\end{equation}
with $J:\RR^x\times \RR^m\to\RR^m$ 
bounded and satisfying
\eqref{eq:Upsilon} and $\epsilon$ is a small enough strictly positive
real number.
\end{list}

Whatever design route a), b) or c) we follow, we obtain the following lemma.
\begin{lemma}
\label{lem1}
Under Assumptions \ref{ass:stability} and \ref{ass:staticgain}, the function $V_e$ is
positive definite and proper on $\Stab\times \RR^r$. Its derivative along the extended
system (\ref{eq:extended}) 
in closed-loop with $u=\psi(x,z)$ is negative definite 
on $\Stab\times \RR^r$ and upperbounded by a negative definite quadratic 
form of $(x,z)$ in a 
neighborhood of the origin. Consequently, for the corresponding closed 
loop system, the origin is asymptotically stable with $\Stab\times\RR^r $
as domain of attraction (without forgetting 
Remark \ref{rem1}) and locally exponentially stable.
\end{lemma}

\begin{IEEEproof}
Since $V$ is positive definite and proper on $\Stab$, $V_e$ is positive definite and proper on 
$\Stab\times\RR^r$. Also the derivative of $V_e$ along the solutions of the 
closed oop system is negative definite in $(x,\psi(x,z))$
and upperbounded by a negative definite quadratic 
form of $(x,\psi(x,z))$ in a 
neighborhood of the origin (see \cite{Mazenc96,Teel96} 
for example). With this, to complete the proof, it is 
sufficient to show the existence of a real number $c$ such that
$$
|z| \leq c\,  |\psi (0,z)|
\  .
$$
Since we have 
\\[0.7em]\null \hfill $\displaystyle 
\psi (0,z) =  J(0,L_gH(0) z)
\; ,
$\hfill respectively \hfill $\displaystyle 
=  J(0,H_0 g(0) z)
$\hfill\null  \\[0.7em]
where the function $J$ satisfies (\ref{eq:Upsilon}), the above 
inequality holds if  $L_gH(0) $, respectively $H_0g(0)$, is right 
invertible. But, by differentiating (\ref{eq:beta}) which 
holds at least in a neighborhood of the origin, using 
(\ref{eq:LP21}) and (\ref{eq:beta0}), and since $f$ and $\beta 
$ are zero at the origin, we have
$$
\frac{\partial H}{\partial x}(0)\;=\; H_0
\  .
$$
Assume the matrix $H_0g(0)$ is not right invertible, i.e. the exists a vector 
$v$ in $\RR^r$ such that
$$
v^\top H_0g(0)\;=\; 0\  .
$$
Then we have
$$
\left(\begin{array}{@{}cc@{}}
v^\top H_0 & - v^\top  \frac{\partial k}{\partial y_r}(0,0)
\end{array}\right)
\left(\begin{array}{@{}cc@{}}
\frac{\partial f}{\partial x}(0) & g(0)
\\[0.5em]
\frac{\partial h_r}{\partial x}(0)  & 0
\end{array} \right)\;=\; 0
$$
which contradicts Assumption \ref{ass:staticgain}.
\end{IEEEproof}

\begin{remark}~
\begin{itemize}
\item
Because the set $\U$ in Assumption \ref{ass:observer} is star-shaped, 
while
satisfying (\ref{eq:Upsilon}), the function $J$ can  
always be chosen such that that the function $\psi$ above defined takes values in $\U$.
\item A drawback of the integral action is the 
possible wind-up. To prevent this phenomenon, in all the above, $\dot 
z$ can be modified in
\begin{equation}
\label{eq:LP3}
\dot z = k(x,y_r) + \omega \left[\texttt{sat}_{\bar z}(z+H(x)) - 
(z+H(x))\right]
\end{equation}
with $H(x)$
replaced by $H_0x$ when needed and
where the saturation function is defined in (\ref{eq:sat}), $\omega $ is any strictly positive real number and
$\bar z$ should be chosen large enough \startmod to allow the $z$-dynamics
to converge to the right equilibrium point. \stopmod This modification does not change 
anything to the asymptotic stability which can be established with 
the same Lyapunov functions.
\end{itemize}
\end{remark}

\subsection{Definitions of $\StabObs  $ and  $\mu$ and saturation of $\psi$ to get the function $\bpsi$}
\label{sec:designProp}
If we were to design a state feedback, we could stop here. But the 
output feedback we design is based on the previous state 
feedback and augmented with an observer. Since the 
estimated state may make no sense during some transient periods, we need a mechanism
to prevent any bad closed-loop effects during these periods. As 
proposed in  \cite{Khalil-Esfandiari}, we use saturation.

First we define the set $\StabObs $ where we would like 
the state to remain. For this, let $\Stab$  be given by Assumption 
\ref{ass:stability}, maybe 
modified as explained in Remark \ref{rem1} above. Similarly, let 
$\Obs $ be given by Assumption 
 \ref{ass:observer} (maybe modified later as in (\ref{eq:LP5})).
Let also the function $V_e$, positive definite and proper on 
$\Stab\times\RR^r$, be given by the above design of the state feedback or a converse Lyapunov 
theorem  \cite{Kurzweil56} satisfying
\\[0.7em]$\displaystyle 
\dot V_e(x,z)
$\hfill \null \\[0.5em]\null\hfill$
\begin{array}[b]{@{}c@{\  }l@{}r@{}}
=&\displaystyle \frac{\partial V_e}{\partial x}(x,z)
\!\left[f(x)+g(x) \psi (x,z)\right]
&+
\frac{\partial V_e}{\partial z}(x,z)k(x,h_r(x))
\\[0.5em]
= &\displaystyle -W_e(x,z) &
\refstepcounter{equation}\label{eq:LP14}(\theequation)
\end{array}
$\\[0.7em]
where the function $W_e$ defined here is
positive definite on $\Stab\times\RR^r $.
Then, if $\Stab$ is not a subset of ${\Obs }$, we let $\vv_\infty$ 
be the real number defined as
\begin{equation}
\label{eq:vinfty}
\vv_\infty =
\displaystyle \inf_{(x,z)\in (\Stab\times\RR^r )\setminus (\Obs  
\times \RR^r)} V_e(x,z)
\  .
\end{equation}
If not let formally $\vv_\infty$ be infinity.
We define the open set\footnote{%
See the further modification (\ref{eq:LP13})
}
\be\label{eq:setSO}
\StabObs  = \{(x,z)\in \Stab\times\RR^r  \; : \; V_e(x,z) < \vv_\infty \}\;.
\ee
This set in non empty since it contains the origin.

In the same way,
to
each
 real number $\vv$ in $[\,0 ,\vv_\infty)$ we associate the set
\begin{equation}
\label{eq:LP10}
\Omega_{\vvs} = \{(x,z)\in \Stab\times\RR^r \; : \; V_e(x,z) \leq  \vv \} \;.
\end{equation}
It is a compact subset of $\StabObs $. Also, from Lemma 
\ref{lem1}, it is forward invariant
for the extended system (\ref{eq:extended}) in closed-loop with 
$u=\psi(x,z)$.
On the other hand, for any $\mathcal{C} _{x\!z}$, compact subset of $\StabObs  $,
we can find real numbers $\vv_1<\vv_2$ satisfying
\be\label{eq:LP7}
\mathcal{C}_{x\!z} \subsetneqq \Omega_{\vvs_1} \subsetneqq 
\Omega_{\vvs_2} \subsetneqq \StabObs  \;.
\ee
Then, with $\mu $ the real number defined  as
\begin{equation}
\label{eq:LP15}
\mu = (1+\s)\max_{(x,z)\in\,\Omega_{\vvs_2}} |\psi (x,z)|
\  ,
\end{equation}\startmod
with $\s$ a small number as in \eqref{eq:sat}, \stopmod
we consider the subset $\U(\mu)\subset \U$ (see \eqref{eq:Uu}). As 
$\U$ in Assumption \ref{ass:observer}, it is star-shaped with the origin as a star-center.
Let then the function $\bpsi:\RR^n\times\RR^r \to \U(\mu)$ be 
\be\label{eq:bpsi}
\bpsi(x,z) = \texttt{sat}_{\mu} (\psi (x,z))
\  .
\ee
It is bounded and Lipschitz and, as $\psi$, it is $C^1$ on a neighborhood of the origin.
Similarly, we modify the function $k$  (defined in \eqref{eq:kcond1}) by saturating its argument $x$. Namely we replace
\\[1em]\null \hfill  $
k(x,h(x))
$\hfill  by  \hfill $k(\texttt{sat}_{\bar x}(x), h_r(x))$
\refstepcounter{equation}\label{eq:integr}\hfill$(\theequation)$
\\[0.7em]
where 
\be\label{eq:satbarx}
\bar x = (1+\s)\max_{(x,z)\in\,\Omega_{\vvs_2}} |x|\,.
\ee

\section{Observer Design }\label{sec:observer}

In the Assumption \ref{ass:observer} we ask for the knowledge of the family of 
observers (\ref{eq:observer}). Fortunately it can be obtained as a
high gain observer. A lot of attention has been devoted to this type of observers and many 
results are available at least for the single output case. See for
example the survey \cite{Khalil-Praly} and the references therein.
We are interested here in some specific aspects as
\begin{list}{}{%
\parskip 0pt plus 0pt minus 0pt%
\topsep 0pt plus 0pt minus 0pt
\parsep 0pt plus 0pt minus 0pt%
\partopsep 0pt plus 0pt minus 0pt%
\itemsep 0pt plus 0pt minus 0pt
\settowidth{\labelwidth}{(a)}%
\setlength{\labelsep}{0.5em}%
\setlength{\leftmargin}{\labelwidth}%
\addtolength{\leftmargin}{\labelsep}%
}
\item[(a)]
the possibility of writing the dynamics of the observer in the 
original coordinates;
\item[(b)]
the multi-output case;
as far as we know at the time we write this text, the study of tunable observers in the multi-output case 
is far from being conclusive. Only some sufficient conditions are known
(see, for instance 
\cite{Besancon07, Khalil-Praly, Tornambe92, HammouriBornard10, Bornard, GK, Atassi-Khalil});
\item[(c)]
the fact that observability holds only on $\Obs$, a (possibly) 
strict subset of the full space $\RR^n$.
\end{list}
To introduce them, we find useful to start with a very 
brief reminder on single output high gain observers.

\subsection{Reminder on high gain observers in the single output case}
\label{sec:sisohgo}

It is known (see \cite[Theorem 3.4.1]{GK} for example) that, for a single-input single-output system of the form
\be\label{eq:sisosys}
\dot x = f(x) + g(x)u \;, \quad y = h(x)\;, \quad x\in \RR^n,\; u, y \in \RR\;,
\ee
which is observable uniformly  with respect to the input and is 
differentially observable of order $n_o$, there exists  an injective 
immersion $\Phi:\RR^n\to\RR^{n_o}$, obtained as 
\be\label{eq:sisophi}
\phi=\Phi(x) = \Big(  h(x) \quad L_f h (x) \quad \cdots \quad L_f^{n_o-1}h(x) \Big)^\top\;,
\ee
which puts the system \eqref{eq:sisosys} into the so called observability (triangular) normal form 
\be\label{eq:sisotriang}
\dot \phi = A_{n_o}\phi + B_{n_o} b(\phi) + D_{n_o}(\phi)u\,, \qquad y = C_{n_o} \phi
\ee
where
\be\label{eq:primeform}
\ba{@{}r@{\: }c@{\: }l}
A_{n_o} & = & \bpmx 0_{{n_o}-1 \times 1} & I_{{n_o}-1\times {n_o}-1} \\ 0 & 0_{1\times {n_o}-1}\epmx,
\   B_{n_o} \: = \: \bpmx 0_{{n_o}-1 \times 1 } \\ 1\epmx,  \\[1.3em]
C_{n_o} & = & \bpmx 1 & 0_{1\times {n_o}-1}\epmx, \\[.5em]
D_{n_o}(\phi) & = & \bpmx d_1(\phi_1), \ldots, d_i(\phi_1, \ldots, \phi_i), \ldots, d_{n_o}(\phi)\epmx^\top
\ea\ee
and where $b(\cdot)$, $d_i(\cdot)$ are locally Lipschitz function.
An observer for the system (\ref{eq:sisosys}) is
\be\label{eq:sisohgo}\ba{rcl}
\dot {\hphi} & = & A_{n_o}\hphi + B_{n_o} b(\hphi) + D_{n_o}(\hphi)u 
\qquad \qquad 
\\
\multicolumn{3}{r@{}}{+ K_{n_o} L_{n_o}(\ell)(y - C_{n_o} \hphi)\,, }
\\[.5em]
\hat x & = & \Phi^{\ell\mbox{\small -}inv}(\hphi)\,,
\ea\ee
where $K_{n_o}$ is such that $(A_{n_o} - K_{n_o} C_{n_o})$ is Hurwitz, 
$L_{n_o}(\ell) = {\rm diag}(\ell, \ldots, \ell^{{n_o}})$ and 
$\Phi^{\ell\mbox{\small -}inv}$ is any locally Lipschitz left inverse function of 
$\Phi$ satisfying
$$
\Phi^{\ell\mbox{\small -}inv}(\Phi(x))\;=\; x\qquad \forall x\in \RR^n
\  .
$$
In the $\phi$-coordinates it is a standard high gain observer \startmod the dynamics
of which \stopmod
can be made arbitrary fast by 
increasing the high-gain parameter $\ell$ (see for instance \cite{Bornard}).

\subsection{On the possibility of writing the dynamic of the observer in the original coordinates}

As already observed in \cite{Maggiore-Passino},
a main issue in implementing the observer (\ref{eq:sisohgo})
is about the function $\Phi^{\ell\mbox{\small -}inv}$ for which we 
have typically no analytical expression, meaning that we have to 
solve on-line \startmod a minimization problem as \stopmod
$$
\hax\;=\; \textsf{argmin}_x |\phi(x)-\hphi|\,.
$$
Fortunately as noticed in \cite{Deza} and proposed also in \cite{Maggiore-Passino},
this difficulty can be rounded when $\Phi$ is a diffeomorphism. 
Indeed in this case $\phi$ is simply another set of coordinates for 
$x$ and the observer (\ref{eq:sisohgo}) can be simply rewritten in the original $x$ 
coordinates as
\be\label{eq:sisohgocoor}
\dot {\hat x} = f(\hat x) + g(\hat x)u + \left( \pderiv \Phi x(\hat x)\right)^{\!\!-1} K_n L_n(\ell)(y - h(\hat x))\,.
\ee
As a consequence there is no need to find the inverse mapping of the 
function $\Phi$ but, \startmod (infinitely) more simply, only \stopmod to invert the matrix $\pderiv \Phi x(\hat x)$.
But for $\Phi$ to be a diffeomorphism, we need $n_o$ to be equal to 
$n$, i.e. to have the (full order) observer to have the smallest 
possible dimension.

\subsection{High gain observer in the multi-output case}
As shown in \cite{Tornambe92},
in the multi-input multi-output case \eqref{eq:system}, a typical
expression for $\Phi$ is
\be\label{eq:mimophi}\ba{rcl}
\Phi(x) & = & \Big( \Phi_1(x) \quad \cdots \quad \Phi_p(x) \Big)^\top \;,\\
\Phi_i (x)& =  & \Big(  h_i(x) \quad L_f h_i (x) \quad \cdots \quad L_f^{p_i}h_i(x) \Big)^\top\;,
\ea\ee
where $h_i$ is the $i$-th component of $h$ and $p_i$ are integers
called the observability indexes and $\sum_{i=1}^{p}p_i \geq  n$. The 
dynamics of system \eqref{eq:system} expressed in these coordinates is
\be\label{eq:mimohgocoord}
\dot \phi = A \phi + B \bar b(\phi) + D(\phi) u \;, \qquad y = C\phi
\ee
where
\[\ba{rcl}
A & = & {\rm blckdiag} \,\big(A_{p_1}, \ldots, A_{p_p}\big)\;, \\[.5em]
B & = & {\rm blckcol} \, \big(B_{p_1}, \ldots, B_{p_p}\big)\;, \\[.5em]
C & = & {\rm blckrow} \, \big(C_{p_1}, \ldots, C_{p_p}\big)\;, \\[.5em]
\bar b(\phi) & = & \big( b_1(\phi), \ldots, b_p(\phi) \big)^\top\;,
\\[.5em]
D(\phi) & = & {\rm blckcol} \, \big(D_{p_1}(\phi ), \ldots, 
D_{p_p}(\phi )\big)
\; ,
\ea\]
where $\bar b(\phi)$ and $D(\phi)$ are locally Lipschitz functions.
However, even when the system is observable uniformly  in the 
input, the functions $\bar b$ and $D$ may not have the triangular structure
we need for the design of a high-gain observer. 
Conditions 
under which we do get triangular dependence for $\bar b(\phi)$ 
and $D(\phi)$ have been studied for instance in \cite{Bornard}  and
\cite{HammouriBornard10}.
Going on along this route and imposing $\Phi$ to be 
a diffeomorphism (in order to write the observer in the original coordinates, as done in \eqref{eq:sisohgocoor}), 
an alternative condition under which we do have an appropriate 
structure is given by the following (technical) assumption, for which 
we do not need to know
the inverse of $\Phi$.

\begin{assumption}
\label{ass:existHGO}
There exist
\begin{list}{}{%
\parskip 0pt plus 0pt minus 0pt%
\topsep 0.5ex plus 0pt minus 0pt%
\parsep 0pt plus 0pt minus 0pt%
\partopsep 0pt plus 0pt minus 0pt%
\itemsep 0.5ex plus 0pt minus 0pt
\settowidth{\labelwidth}{i)}%
\setlength{\labelsep}{0.5em}%
\setlength{\leftmargin}{\labelwidth}%
\addtolength{\leftmargin}{\labelsep}%
}
\item[i)]
an open  set $\Obs \subset\RR^n$ containing the origin and a
star-shaped set $\U$ with the origin as star-center,
\item[ii)]
a $C^1$ function $\Phi:\Obs  \to \RR^n$,
\item[iii)]
sequences of matrices
 $L_\ell\in\RR^{n\times n}$,
$M_\ell \in\RR^{n\times n}$ and $N_\ell \in\RR^{p\times p}$,
a matrix $C \in \RR^{p\times n}$, 
\item[iv)]
matrix functions
 $u\in \U\mapsto K(u) \in\RR^{n\times p}$ 
and $u\in \U \mapsto A(u) \in \RR^{n\times n}$,
\item[v)]
and,  for any positive real number $\bar u$, there exist
a positive definite symmetric matrix $P \in \RR^{n\times n}$ and
strictly positive real numbers $\nu$ and $d$,
\end{list}
such that
\begin{list}{}{%
\parskip 0pt plus 0pt minus 0pt%
\topsep 0.5ex plus 0pt minus 0pt%
\parsep 0pt plus 0pt minus 0pt%
\partopsep 0pt plus 0pt minus 0pt%
\itemsep 0.5ex plus 0pt minus 0pt
\settowidth{\labelwidth}{O1) }%
\setlength{\labelsep}{0.5em}%
\setlength{\leftmargin}{\labelwidth}%
\addtolength{\leftmargin}{\labelsep}%
}
\item[O1) ]
the function $\Phi$ is a diffeomorphism on the set $\Obs $
and $\Phi(0)=0$\;,
\item[O2) ]
$\displaystyle
\displaystyle C\,  \Phi(x)\;=\; h(x)\;,
$\hfill \null 
\item[O3) ]
the matrices $A(u),K(u),P,C$ satisfy, for any $u\in \U(\bar u)$,
\begin{eqnarray*}
&\hskip -0.5em\displaystyle P(A(u)-K(u)C)+ (A(u)-K(u)C)^\top P 
\leq  -
2   \nu   P
\,  ,
\\
&\displaystyle
A(u) \,  L_\ell\;=\; L_\ell \,  M_\ell \,  A(u)
\  ,\quad
N_\ell\,  C \,  L_\ell\;=\; C\  ,
\end{eqnarray*}
\item[O4) ]
the
matrix
 $M_\ell$
is such that $M_\ell P^{-1}$ is symmetric  and
satisfies
$$
\lim_{\ell \to +\infty } \lambda _{\min}(M_\ell
P^{-1})\;=\; +\infty\;,
$$
\item[O5) ]
$\displaystyle
\lambda_{\max{}}\left(L_\ell
M_\ell P^{-1}
 L_\ell^\top\right)
\, \leq \, \lambda_{\min{}} (M_\ell
P^{-1}
)^d\;, $  \hfill \null
\\[0.7em]
$\displaystyle
1 \, \leq \,
\lambda_{\min{}}\left(L_\ell
M_\ell P^{-1}
 L_\ell^\top\right)\,
\lambda_{\min{}}(M_\ell
P^{-1}
)^d
\,.
$ \hfill \null
\end{list}
Moreover, for any compact set $\mathfrak{C}$ and
$\widehat{\mathfrak{C}}$ satisfying
$$
\mathfrak{C}\; \subset \; \widehat{\mathfrak{C}}\; \subset\; \Obs 
$$
there exists a sequence of positive real numbers $c_\ell$ such that
\begin{list}{}{%
\parskip 0pt plus 0pt minus 0pt%
\topsep 0.5ex plus 0pt minus 0pt%
\parsep 0pt plus 0pt minus 0pt%
\partopsep 0pt plus 0pt minus 0pt%
\itemsep 0.5ex plus 0pt minus 0pt
\settowidth{\labelwidth}{O1) }%
\setlength{\labelsep}{0.5em}%
\setlength{\leftmargin}{\labelwidth}%
\addtolength{\leftmargin}{\labelsep}%
}
\item[O6) ]
$\displaystyle
\lim_{\ell\to +\infty } c_\ell\;=\; 0\  ,
$\hfill \null 

\item[O7) ]
the function $B:\RR^{n\times m}\to \RR^n$ defined as
\begin{equation}
\label{eq:LP1}
B(\Phi(x),u) =L_f\Phi(x) + L_g \Phi(x)u- A(u)\,\Phi(x)\;,
\end{equation}
satisfies, for all $x_a \in \mathfrak{C}$, $x_b \in \widehat{\mathfrak{C}}$ and $u \in \U(\bar u)$,
\begin{multline}\label{eq:Bbounded}
\left|P^{\frac12} M_\ell^{-1}L_\ell^{-1} 
\left[B(\Phi(x_a),u)-B(\Phi(x_b),u)\right]\right|
\\
\leq \;c_\ell\,  \left|
P^{\frac12}
L_\ell^{-1} \left[\Phi (x_a)-\Phi(x_b)\right]\right|
\  .
\end{multline}
\end{list}
\end{assumption}

\begin{remark}~
\begin{itemize}
\item As shown in the next Lemma, the existence of a high-gain observer for the system 
\eqref{eq:system} is guaranteed if Assumption \ref{ass:existHGO} holds.
In particular the properties O1, O2, O3, O6 and O7 guarantee
 the existence of a converging observer in the original coordinates whereas
properties O4 and O5 assure
its tunability property. 

\item We remark that these conditions can be checked without need of finding formally the inverse mapping $\Phi^{-1}$. 
In particular, given a system and a candidate diffeomorphism $\Phi$ (property O1), 
one can immediately check properties O2 (linear dependence of the diffeomorphism on the output) 
Then, if this properties holds, one 
can fix the degrees of freedom $K(u)$, $M_\ell$, $N_\ell$, $L_\ell$, $P$ which properly defines the high-gain 
observer as shown later in Lemma \ref{lem:observerR} (see \eqref{eq:LP8})
\startmod and check also the Lipschitz
condition \eqref{eq:Bbounded} in 07. 
Finally,  property O3 guarantees the convergence of the observer (see proof of Lemma \ref{lem:observerR}). \stopmod

\item 
The conditions of  Assumption \ref{ass:existHGO}
are satisfied in the single-output case considered
 in Section \ref{sec:sisohgo} when $n_o = n$,
by choosing $\Phi$ as in \eqref{eq:sisophi},
and picking
\[
L_\ell  =  \text{diag }(1,\ell,\ldots,\ell^{n-1})
\  ,\quad
M_\ell  = \ell
\  ,\quad
N_\ell =1
\]
\[
A(u) = A_n, \quad B(\Phi(x),u) = B_n b(\Phi(x)) +  D_n(\Phi(x))u,
\] 
and $C = C_n$, where the triplet $A_n$, $B_n$, $C_n$ and the functions $b(\cdot)$, $D(\cdot)$ are given in \eqref{eq:primeform}. 
In this case, we set $L_n(\ell)  =  L_\ell M_\ell N_\ell$ and $K(u) = 
K_n$ in the observer \eqref{eq:sisohgo}.

\item 
In this assumption $A$ is allowed to be input-dependent to allow a 
broader class of nonlinear systems. For instance it can be verified that the system
\[
\dot x_1 = x_2\,, \qquad \dot x_2 = u\,, \qquad y = -x_1 + x_2+ x_2^2\;,
\]
can not be transformed in the form \eqref{eq:sisotriang} but it satisfies Assumption \ref{ass:existHGO}.
\item
In some cases, the nonlinear terms \eqref{eq:LP1} can be disregarded in 
the high gain observer \startmod design (usually also \stopmod called dirty derivative observer). 
This is possible for example
when the notions of observability indexes and relative degree indexes coincide (see \cite{Seshagiri-Khalil05}
among others).
In this case, these nonlinear terms act through their bound 
and not their Lipschitzness. Unfortunately then  \startmod
a very specific structure is needed because otherwise
the gain between these nonlinear terms \stopmod and some estimation error is increasing with the observer gain.
Here we intend to \startmod consider \stopmod a broader class of systems and thus we do 
need to have these terms present in the observer.
\end{itemize}
\end{remark}

\begin{lemma}\label{lem:observerR}
Under Assumption \ref{ass:existHGO}, for any compact set $\mathfrak{C}$ and
$\widehat{\mathfrak{C}}$ satisfying
$$
\mathfrak{C}\; \subset \; \widehat{\mathfrak{C}}\; \subset\; \Obs 
\  ,
$$
the family of systems
\begin{equation}
\label{eq:LP8}
\dot \hax_\ell = f (\hax_\ell) +g(\hax_\ell)u 
+\left(\pderiv \Phi x (\hax_\ell)\right)^{-1}  \!\!\!\!\! L_\ell M_\ell K(u) N_\ell  \big[y-h(\hax_\ell)\big]  
\end{equation}
indexed by $\ell$ in $\RR_{>0}$ satisfies points 2 to 6 of Assumption 
\ref{ass:observer}.
\end{lemma}

\begin{IEEEproof}
We let
\begin{equation}
\label{eq:LP18}
\phi  = \Phi(x)\;,\qquad \hphi_\ell = \Phi(\hax_\ell)\;, \qquad \tphi_\ell =
\phi -\hphi_\ell\;.
\end{equation}
With (\ref{eq:LP1}) and
(\ref{eq:LP8}), \startmod systems \eqref{eq:system} and \eqref{eq:LP8} are transformed in \stopmod
\[\begin{split}
\dot \phi  & = A(u)\phi  + B(\phi,u )\\
\dot \hphi_\ell & = A(u)\hphi_\ell + B(\hphi_\ell,u)+L_\ell M_\ell K(u)N_\ell C(\phi -\hphi_\ell)
\end{split}\]
With Assumption \ref{ass:existHGO} and the notations (\ref{eq:LP18}), we
define the Lyapunov Function
$$
U_\ell (x,\hax)=
\frac{1}{2}
(\phi  - \hphi_\ell)^\top[L_\ell M_\ell P^{-1} L_\ell^\top]^{-1}(\phi -\hphi_\ell)\;.
$$
As $\Phi$, it is defined on $\Obs \times{\Obs }$ and it takes values in
$\RR_{\geq 0}$. Also, because the matrix $L_\ell M_\ell P^{-1} L_\ell^\top$
is positive definite, we have
$$
\forall (x,\hax_\ell)\in\Obs \times\Obs \,, \quad U_\ell(x,\hax_\ell)=0
\quad \Longleftrightarrow \quad x=\hax_\ell
\  .
$$
So point 2 of Assumption \ref{ass:observer} holds.
Also we get
\\[0.5em]$\displaystyle 
\dot U _\ell (x,\hax) = (\phi -\hphi_\ell)^\top\!\!
\left[L_\ell^{-\top}P M_\ell^{-1} L_\ell^{-1}\right]\!\!\times
$\refstepcounter{equation}\label{eq:LP9}\hfill$(\theequation)$
\\[0.3em]\null\hfill$
\times \left[
\vrule height 0.6em depth 0.6em width 0pt
(A(u)-L_\ell M_\ell K(u)N_\ell C) (\phi -\hphi_\ell) +
B
(\phi,u )-
B
(\hphi_\ell,u) \right]
$ \\[0.5em]
which, with using O3 and \eqref{eq:Bbounded}, gives,
for all $(x,\hax)$ in $\mathfrak{C}\times \widehat{\mathfrak{C}}$,
$$
\dot U _\ell \leq - \nu |
P^{\frac12}
L_\ell^{-1}\tphi_\ell|^2 + c_\ell |
P^{\frac12}
L_\ell^{-1}\tphi_\ell|^2 (1+|u|)
\;.
$$
Then,
with O6,
for any $|u|\leq\bar u$, there exists a $\underline{\ell}$ such that, for any $\ell \geq \underline\ell$,
\be\label{eq:LP4}
\dot U_\ell(x,\hax) \leq -\frac\nu2 \tphi_\ell^\top L_\ell^{-\top} P L_\ell^{-1}\tphi_\ell
\qquad \forall (x,\hax)\in \mathfrak{C}\times \widehat{\mathfrak{C}}\;.
\ee
Since we have
$$
P\; \geq \; \lambda _{\min}(P) \lambda _{\min}(M_\ell P^{-1})\,  P M_\ell
^{-1}
\  ,
$$
we obtain, for all $(x,\hax)$ in $\mathfrak{C}\times 
\widehat{\mathfrak{C}}$,
$$
\dot U_\ell(x,\hax) \leq - \;
\frac{\nu \,\lambda _{\min}(P) \lambda _{\min}(M_\ell P^{-1})}{2}
\; U_\ell(x,\hax)
\  .
$$

So, with O4, points 4 and 5 of Assumption \ref{ass:observer}
hold when we
choose the integer $\ellmod$ as the integer part of the ratio
$\ell/\underline{\ell}$ and with
$$
\sigma_\ellmod =
\frac{\nu \,\lambda _{\min}(P) \lambda _{\min}(M_\ell P^{-1})}{2}
\  .
$$

Next, we have
\vspace{-2ex}
\begin{multline*}
U_\ell (x,\hax) \lambda_{\min{}}(L_\ell M_\ell P^{-1} L_\ell^\top)
=
\dfrac{\tphi_\ell^\top \;(L_\ell M_\ell P^{-1} L_\ell^\top)^{-1}\; \tphi_\ell}
{\lambda_{\max}((L_\ell M_\ell P^{-1} L_\ell^\top)^{-1})}  
\\ \leq 
|\phi -\hphi_\ell|^2 
\end{multline*}
\vspace{-5ex}
\begin{multline*}
|\phi -\hphi_\ell|^2 
 \leq  
\dfrac{\tphi_\ell^\top \; (L_\ell M_\ell P^{-1} L_\ell^\top)^{-1} \; \tphi_\ell}
{\lambda_{\min}((L_\ell M_\ell P^{-1} L_\ell^\top)^{-1})} 
\\
 \leq 
 U_\ell(x,\hax) \; \lambda_{\max{}}(L_\ell M_\ell
P^{-1}
L_\ell^\top)
\end{multline*}
So, with O5, we get
\\[0.5em]$\displaystyle 
U_\ell(x,\hax) \lambda_{\min{}}( M_\ell
P^{-1}
)^{-d}
$\hfill\null \\\null \hfill  $\displaystyle \leq 
|\Phi(x)-\Phi(\hax_\ell)|^2  \leq  U_\ell(x,\hax)   \lambda_{\min{}}( M_\ell
P^{-1}
)^d.
$\\[0.5em]
But, because $\Phi$ is a diffeomorphism defined on $\Obs $, for
any compact subsets $\mathfrak{C}$ and $\widehat{\mathfrak{C}}$ of
$\Obs $, there exist real numbers $\overline{\Phi }$ and
$L_{\Phi^{-1}}$, independent of $\ell$, such that, for all $x$ in $\mathfrak{C}$ and $\hax_\ell$
in $\widehat{\mathfrak{C}}$, we have
\vspace{-2ex}
\begin{multline*}
|x-\hax_\ell| \; = \;
|\Phi^{-1}(\Phi(x))-\Phi^{-1}(\Phi(\hax_\ell))|
\\ \leq \; 
L_{\Phi^{-1}}
|\Phi(x)-\Phi(\hax_\ell)|
\; \leq \; \overline{\Phi }
\  .
\end{multline*}
This gives
\\[0.5em]$\displaystyle 
|x-\hax_\ell|^2 \frac{1}{L_{\Phi^{-1}}^2}\lambda_{\min{}}( M_\ell
P^{-1}
)^{-d} \leq \, U_\ell(x,\hax)  \,
$\hfill \null \\\null \hfill $\displaystyle \leq \,
\overline{\Phi }^2 \lambda_{\min{}}( M_\ell
P^{-1}
)^{d}.
$\\[0.5em]
So, with O4,
point 3 of Assumption \ref{ass:observer}
holds.

Finally, point 6 of Assumption \ref{ass:observer}
holds too. Indeed, by definition of the set $\U(\bar u)$, the matrices $K(u), M_\ell, 
N_\ell, L_\ell$
and the diffeomorphism $\Phi$, there exists a positive definite function $L_{\vartheta_\ell}(\hax_\ell)$
such that
\[
\left| 
\left(\pderiv \Phi x(\hax_\ell) \right)^{-1}
L_\ell M_\ell K(u) N_\ell
\, \right| 
\; \leq \; 
L_{\vartheta_\ell}(\hax_\ell)
\]
for any $\ell>0$, $u\in \U(\bar u)$ and $\hax_\ell\in \widehat{\mathfrak{C}}$.
\end{IEEEproof}

\subsection{Taking care of observability restricted to $\Obs$ by an 
observer modification}
\label{sec:proofobserver}
In the above (\ref{lem:observerR}), we are missing point 1 of Assumption 
\ref{ass:observer}, namely $\widehat{\mathfrak{C}}$ may not be 
forward invariant. The problem is that
the observer \eqref{eq:extendedobserver} does not guarantee that
$\hax_\ell$ remains in $\Obs $ and therefore that $\frac{\partial
\Phi}{\partial x}(\hax_\ell)$ is invertible. To round this problem, as in
\cite{Maggiore-Passino},
we modify this observer, here not by projection, but by considering a
dummy measured output (extending the results in \cite{AstolfiPraly12}).
To make our point clear, we introduce the following assumption.

\begin{assumption}\label{ass:h2}
Given the set $\Obs $ and the diffeomorphism $\Phi$ of Assumption 
\ref{ass:existHGO}, for any compact subset $\mathfrak{C}$ of $\Obs $,
we know of a $C^1$ function
$h_2:\Obs \to \RR_{\geq 0}$ such that:
\begin{list}{}{%
\parskip 0pt plus 0pt minus 0pt%
\topsep 0.5ex plus 0pt minus 0pt%
\parsep 0pt plus 0pt minus 0pt%
\partopsep 0pt plus 0pt minus 0pt%
\itemsep 0.5ex plus 0pt minus 0pt
\settowidth{\labelwidth}{H1.}%
\setlength{\labelsep}{0.5em}%
\setlength{\leftmargin}{\labelwidth}%
\addtolength{\leftmargin}{\labelsep}%
}
\item[H1.]
the set
$
\left\{x\in \RR^n\,  :\: h_2(x)< 1\right\}$
is a subset of $\Obs $;
\item[H2.]
the function
$x\mapsto
\frac{h_2(x)}{\left|\frac{\partial h_2}{\partial x}(x)\right|}$ is continuous on
$\Obs $;
\item[H3.]
for any real number $s$ in $[0,1]$, and any $x_1$ and
$x_2$ in $\Obs $  satisfying
$$
h_2(x_1)\leq s
\quad ,\qquad
h_2(x_2)\leq s\;,
$$
we have $ \quad h_2(x)\leq s \quad $ 
for all $x$ which satisfies for some $\lambda $ in $[0,1]$
$$
\Phi(x)\;=\; \lambda \Phi(x_1) + (1-\lambda ) \Phi(x_2)
\  .
$$
This means nothing but the fact that,
for any $s$ in $[0,1]$,
the image by $\Phi$ of the set
$\{x\in \RR^n:\,  h_2(x) \leq s\}$
is convex;
\item[H4.]
the set $\Obs _{mod}$ defined as
\begin{equation}
\label{eq:LP5}
\Obs _{mod}= \{x\in \RR^n:\,
h_2(x)
\leq
 0\}
\end{equation}
contains $\mathfrak{C}$ and
has a non empty interior which contains the origin;
\item[H5.]
the set
$$
\widehat{\mathfrak{C}}\;=\; \left\{x \in \RR^n \;: \;|h_2(x)| \leq
\textstyle \frac12 \right\}
$$
is compact.
\end{list}
\end{assumption}

\begin{remark}~
\begin{itemize}
\item
There is a systematic way to define this function $h_2$ when,
given the compact set $\mathfrak{C}$, we know
 a positive definite symmetric matrix $Q$ and a real
number $R$ satisfying
$$
\Phi(\mathfrak{C})\; \subset\;
\{\phi \in\RR^n: \phi ^\top Q\phi \leq R\}
\; \subset \; \Phi(\Obs )
\  .
$$
Indeed, in this case we let $\varrho$ be the
number defined as
$$
\varrho\;=\;
\sup_{R:\{\phi : \phi ^\top Q\phi \leq R\}\subset \Phi(\Obs )} R \  .
$$
Since $\Obs $ is a neighborhood of the origin, $\varrho$ is
strictly positive.
Then we select a real number $\epsilon$
in $(0,1)$ and let
\be\label{eq:h2}
h_2(x)\;=\; \max\left\{\frac{\Phi(x) ^\top Q \Phi(x)}{\varrho} - \epsilon
\,  ,\,  0\right\}^2
\  .
\ee
With this choice and since $\Phi$ is a diffeomorphism, we can check that Properties H1 to
H5 are satisfied.
\item
We may dislike the convexity property mentioned in H3 of Assumption 
\ref{ass:h2}.
Unfortunately it is in some sense necessary. Indeed our objective
with the modification $E$ is to preserve the high-gain paradigm.
This means in particular that we choose to keep an Euclidean distance
in the image by $\Phi$ as a Lyapunov function for
studying the error dynamics.
Also
we need an infinite gain margin, as defined in Definition 2.8 in
\cite{Sanfelice.Praly.12}, since the correction term must dominate all the other ones in the expression of $\dot \hax$
when $h_2$ becomes too large. Then as proved in Lemma 2.7
\cite{Sanfelice.Praly.12}, with such constraints, the convexity
assumption is necessary.
This implies that, if we want to remove the convexity assumption, we
have to find another class of observers.
\end{itemize}
\end{remark}

We are interested in the function $h_2$ because it satisfies the property
$$
h_2(x) = 0\qquad \forall x\in \Obs _{mod}
\; .
$$
This leads us to introduce a dummy measured output
$$
y_2=h_2(x)
\  .
$$
Indeed $y_2$  is zero when $x$ is in $\Obs _{mod}$. But 
$\Obs _{mod}$ being a strict subset of $\Obs $, we have 
here a stronger constraint.
To deal with this restriction, we need to ``reduce'' the set $\StabObs $ by
modifying its definition given in \eqref{eq:setSO} into
\begin{equation}
\label{eq:LP13}
\begin{array}{rcl}
\vv_\infty &=&\displaystyle
\inf_{(x,z)\in(\Stab\times\RR^r) \backslash (\Obs _{mod} \times \RR^r)} V(x,z)
\\[1em]
\StabObs  &=&\big\{(x,z)\in (\Stab\times\RR^r) \; : \; V(x,z) < \vv_\infty \big\}\;.
\end{array}
\end{equation}

With Assumption \ref{ass:h2}, point 1 of Assumption \ref{ass:observer} can be established via 
a modification of the observer.

\begin{lemma}
\label{prop:observer}
Assume Assumptions \ref{ass:h2} holds.
Let $\Phi:\Obs  \to \RR^n$ be a diffeomorphism, $\bar u$ be a positive real number and
$t\to u(t)$ be a continuous function with values in $U(\bar u)$ and $t\to y(t)$ be a continuous bounded function.
The set $\widehat{\mathfrak{C}}$ given in H5 is forward invariant for any system 
in the family, indexed by $\ell$ in $\RR_{>0}$,
\\[0.7em]$\displaystyle 
\dot \hax_\ell = f (\hax_\ell) +g(\hax_\ell)u 
$\refstepcounter{equation}\label{eq:extendedobserver}\hfill$(\theequation)$
\\[0.3em]\null\hfill$
+\left(\pderiv \Phi x (\hax_\ell)\right)^{-1}  \!\!\!\!\! L_\ell M_\ell K(u) N_\ell  \big[y-h(\hax_\ell)\big]  
+
E(\hax_\ell,u,y)
$\\[0.7em]
where the term $E$ is defined as
\\[0.7em]$\displaystyle 
E(\hax_\ell ,u,y) =
-\tau_\ell \, (\hax_\ell,u,y)
\times 
$\refstepcounter{equation}\label{eq:CorrectionTerm}\hfill$(\theequation)$
\\[0.3em]\null\hfill$
\times 
\left(\pderiv \Phi x (\hax_\ell)\right)^{\!\!-1} \!\!\!\!
L_\ell M_\ell P^{-1} L_\ell^\top \!\left(\pderiv \Phi x 
(\hax_\ell)\right)^{\!\!\!-1\top} \!\!
 \pderiv{h_2}{x}(\hax_\ell )^\top h_2(\hax_\ell )
$\\[0.7em]
where $\tau_\ell$ is a $C^1$ function to be chosen large 
(see\eqref{eq:LP6}).

If all conditions of Assumption \ref{ass:existHGO} hold and the model state $x$ remains in 
$\Obs _{mod}$, then all the points of Assumption 
\ref{ass:observer} are satisfied.
\end{lemma}

\begin{IEEEproof}
First we observe that
\\[0.7em]$\displaystyle
\frac{\partial h_2}{\partial x}(\hax_\ell)\dot\hax_\ell  = 
\;R(\hax_\ell,u,y) +
$\hfill \null \\\null \hfill $\displaystyle
- \; \tau_\ell (\hax_\ell)
{\left|
(M_\ell P^{-1})^{-\frac12}
L_\ell^\top
\left(\frac{\partial \Phi}{\partial  x }(\hax_\ell)\right)^{-1\top}
\frac{\partial h_2}{\partial x}(\hax_\ell )^\top
\right|}^2
h_2(\hax_\ell )
$\\[0.7em]
where we have let
\\[0.7em]$\displaystyle
R(\hax_\ell,u,y) =
\frac{\partial h_2}{\partial x}(\hax_\ell) \times
$\hfill \null \\\null \hfill $\displaystyle
\times
\Bigg[f (\hax_\ell) + \,g(\hax_\ell)u
+\left(\pderiv \Phi x (\hax_\ell)\right)^{-1} \!\!\!\!\! L_\ell M_\ell K(u) N_\ell  [y-h(\hax_\ell)]\Bigg]\,.
$\\[0.7em]
This motivates us for choosing $\tau_\ell $ satisfying
\begin{equation}
\label{eq:LP6}
\tau_\ell (\hax_\ell,u,y)\geq
\frac{
8
h_2(\hax_\ell )^2
\: R(\hax_\ell,u,y)
}{
{\left|
(M_\ell P^{-1})^{-\frac12}
L_\ell^\top
\left(\pderiv \Phi x (\hax_\ell)\right)^{-1\top}
\dfrac{\partial h_2}{\partial x}(\hax_\ell )^\top \right|}^{\,2}
}\  
\end{equation}
which can be computed on-line.

Thanks to H2, the function $x \mapsto \tau_\ell (x)$ defined this
way is continuous on $\Obs $.
So we can use $\tau_\ell$ as long as $\hax_\ell$ is in
$\Obs $. It implies that
$\dot {\overparen{h_2(\protect\widehat{x}_\ell )}}$ is
non positive
 when
$h_2(\hax_\ell)$ is  strictly larger than
$\frac{1}{2}$.
With uniqueness of solutions, this implies
that, for each $s$ in $[
\textstyle\frac{1}{2}
,1]$
the set $\{(\hax_\ell):\,  h_2(\hax_\ell) \leq s\}$ is forward
invariant
and so is the compact set $\widehat{\mathfrak{C}}$ in
particular. This says that point 1 of Assumption \ref{ass:observer} hold.

On the other hand, the modification $E$ augments $\dot U_\ell$ in (\ref{eq:LP9}) with
$$
-\tau_\ell (\hax_\ell)  \left[\Phi(\hax_\ell )-\Phi(x)\right]^\top
\left(\pderiv \Phi x (\hax_\ell)\right)^{-1\top}
\frac{\partial h_2}{\partial x}(\hax_\ell )^\top h_2(\hax_\ell )\;.
$$
But, when $h_2(x)$ is zero which is the case when the model state $x$ remains in 
$\Obs _{mod}$ and when $h_2(\hax_\ell)$
is in $[0,1]$, the convexity property of $h_2$ in H3 gives
$$
0\; \leq \; [\Phi (\hax_\ell)-\Phi (x)]^\top
\left(\frac{\partial \Phi}{\partial x}(\hax_\ell)\right)^{-1\top}
\frac{\partial  h_2}{\partial x}(\hax_\ell )^\top h_2(\hax_\ell )
\; .
$$
We conclude that, when all conditions of Assumption 
\ref{ass:existHGO} hold, \eqref{eq:LP4} holds even with the modification $E$.
Hence, from the proof of Lemma \ref{lem:observerR}, points 2 to 5 of Assumption \ref{ass:observer} hold.
Finally,  with (\ref{eq:CorrectionTerm}) and
(\ref{eq:LP6}), the function defined by the right hand side of (\ref{eq:extendedobserver}) 
satisfies also the point 6 of the Assumption \ref{ass:observer}.
\end{IEEEproof}

\begin{remark}
An important feature is that, thanks to the additional term $E$, no
other modification (as saturation) is needed. This
modification, in fact, guarantees that the estimate state $\hax_\ell$
remains in a compact subset of $\Obs $
which depends on the choice of the parameters.
\end{remark}

\section{Proofs of Propositions}
\label{sec:proofs}

\IfJournal{}{
\subsection{Proof of Proposition \ref{prop:outF}}
\label{sec:proofOutF}
This proof 
follows the same 
lines as in \cite{AstolfiPraly12}, inspired by \cite[Chapter 
12.3]{Isidori2} with no meaningful no originality. We give 
it only for the sake of completeness.

We introduce the notations
\[
x_e = \bpmx x\\ z\epmx , \; u_e = \bpmx u_1 \\ u_2 \epmx, 
\;
f_e(x_e,u_e) = \bpmx f(x)+g(x) u_1 \\ k(u_2, h_r(x))\epmx, 
\]
\[
\psi_e(x, z) = \bpmx \psi(x,z) \\ x\epmx,
\quad
\bpsi_e(\hax, z) = \bpmx \bpsi(\hax,z) \\ \texttt{sat}_{\bar x}(\hax)\epmx.
\]
where the functions $\psi$, $\bpsi$ and $\texttt{sat}_{\bar x}$ are defined in Sections \ref{sec2} and \ref{sec:designProp}.
The closed-loop system \eqref{eq:system}, \eqref{eq:controller} can be compactly written as 
$$
\dot x_e = f_e(x_e,u_e)\;, \qquad u_e = \psi_e(\hax, z)\,.
$$
We have a function $V_e$, positive definite and proper on 
$\Stab\times\RR^r$ which allows us to define the compact sets $\Omega _{{{}}\vvs_1}$ and $ \Omega_{{{}}\vvs_2}$,
as in (\ref{eq:LP10}) and satisfying \eqref{eq:LP7}. Also, with 
(\ref{eq:LP14}) and (\ref{eq:bpsi}),  the function
\\[0.7em]$\displaystyle 
-W_e(x,z) \;=\; \dot V_e(x,z)
$\hfill \null \\[0.5em]\null\hfill$
\begin{array}[b]{@{}c@{\  }l@{}}
=&\displaystyle \frac{\partial V_e}{\partial x}(x,z)
\!\left[f(x)+g(x) \bpsi (x,z)\right]
+
\frac{\partial V_e}{\partial z}(x,z)k(x,h_r(x))
\end{array}
$\\[0.7em]
is continuous and negative definite on $ \Omega_{{{}}\vvs_2}$.

These properties imply the following:
\begin{list}{}{%
\parskip 0pt plus 0pt minus 0pt%
\topsep 0.5ex plus 0pt minus 0pt%
\parsep 0pt plus 0pt minus 0pt%
\partopsep 0pt plus 0pt minus 0pt%
\itemsep 0.5ex plus 0pt minus 0pt
\setlength{\labelwidth}{10pt}%
\setlength{\labelsep}{0pt}%
\setlength{\leftmargin}{\labelwidth}%
\addtolength{\leftmargin}{\labelsep}%
}
\item[1. ]
there exists a positive real number $\overline{W}$ and a continuous function $\alpha :\RR_{\geq 0} \to
[0,\overline{W}]$ such that
\\[0.7em]$\displaystyle 
\pderiv{V_e}{x_e}(x_e)\!\left[
f_e(x_e,\bpsi_e(\hax,z))-f_e(x_e,\psi_e (x,z))
\right]
$\hfill \null \\\null \hfill $\displaystyle 
\leq
\alpha (|\hax - x|) \qquad \forall (x,z)\in\Omega_{{{}}\vvs_2}
\: ,\; \forall \hax \in \RR^n\; ;
$
\item[2. ]
there exists a strictly positive real number $\underline{W}$ such
that
\\[0.7em]$\displaystyle
\pderiv{V_e}{x_e}(x_e)
f_e(x_e,\bpsi_e(\hax,z))\; \leq \;
-\underline{W}
$\hfill \null \\\null \hfill $\displaystyle 
\forall (\hax,x,z) : (x,z)\in\Omega_{{{}}\vvs_2}\backslash \Omega_{{{}}\vvs_1}
,\,
|\hax-x|\leq \delta _{xw} ,
$\\[0.7em]
where $\delta _{xw}$ is the strictly
positive real number defined as
$$
\delta _{xw}=
\alpha  ^{-1}\left(
\frac{1}{2}
\min_{(x,z) \in \,\Omega_{{{}}\vvs_2}\backslash \Omega_{{{}}\vvs_1}} W_e(x,z)\right)
\  .
$$
\end{list}
By collecting this, we obtain\\[.5em]
$\pderiv {V_e}{x_e}(x_e)f_e(x_e,\bpsi_e(\hax,z)) $ 	\\[-1em]
\begin{multline}
\label{inequalityWe} 
\leq -W_e(x_e)+\alpha (|\hax-x|)\; \leq \; \overline{W}
\\
\forall \; (x,z) \in
\Omega_{{{}}\vvs_2}\; ,\  \forall \;\hax \in \RR^n,
\end{multline}
\\[-4em]
\begin{multline}\label{inequalityWeconstant}
\leq -\underline{W}
\\
\forall \; (x,z)\in \Omega_{{{}}\vvs_2}\backslash\Omega_{{{}}\vvs_1}
\; ,\
\forall \; \hax  : |\hax-x|\leq \delta _{xw} \;.
\end{multline}

On another hand, let $\mathfrak{C}= (\Omega_{\vvs_2})_x$.
From (\ref{eq:LP10}), it is a compact subset of $\Obs $. With 
this set and $\mu$ defined in (\ref{eq:LP15}) we can invoke
Assumption \ref{ass:observer}. It gives us in particular
the sequence $\sigma _\ellmod$, the integer $d$, the real number
$\overline{U}$ and the function $\alphalow$. This allows us to
define the integer $\underline{\ellmod}$ as the smallest one
satisfying
\begin{equation}
\label{eq:LP16}
\sigma_\ellmod^{2d} \exp\left(-\sigma_\ellmod \frac{\vv_2-\vv_1}{2
\overline{W}}\right)\; \overline{U}
\leq \alphalow \left(\delta_{xw}\right)
\qquad \forall \ellmod\geq \underline{\ellmod}\; .
\end{equation}
From now on, we fix $\ellmod $, arbitrarily but larger than
$\underline{\ellmod}$.

Assumption  \ref{ass:observer}
 gives us also the functions
$\vartheta_\ellmod$ and $U_\ellmod$ and the
forward invariant
compact subset
$\widehat{\mathfrak{C}}$ of $\Obs $. 
Then, because $U_\ellmod$ is continuous and satisfies
the point 2 of Assumption 
\ref{ass:observer},
there exists\footnote{%
\, $\alphaup_\ellmod$ can be constructed from
$s\mapsto \max_{(x,\hax)\in \,\mathfrak{C}\times\widehat{\mathfrak{C}}\,:\,|x-\hax|\leq s} U(x,\hax)$\,.
}
a class-$\mathcal{K}^\infty$ function
$\alphaup_\ellmod$
satisfying
\be\label{inequalityU}
U_\ellmod (x,\hax) \leq \alphaup_\ellmod(|x-\hax|)
\qquad \forall (x,\hax)\in \,\mathfrak{C}\times\widehat{\mathfrak{C}}
\;.
\ee

From $\widehat{\mathfrak{C}}$ we define $\mathcal{C}_{\hax}$ and 
$\Gamma$ as the sets
\be \label{eq:defChatx}
\mathcal{C}_{\hax}= \widehat{\mathfrak{C}}\quad ,\qquad
\Gamma =\widehat{\mathfrak{C}}\times\Omega_{\vvs_2}
\  .
\ee
$\mathcal{C}_{\hax}$ is a compact subset of $\Obs $ and
$\Gamma $ is a compact subset of $\Obs  \times \Omega_{\vvs_2}$. Since all the functions
are
Lipschitz on $\Gamma $, the
solutions of the closed-loop system \eqref{eq:system}, \eqref{eq:controller}
are well defined and unique as long as they are in
the
interior set $\intGamma$ of $\Gamma $.
Moreover
their values satisfy
inequalities \eqref{inequalityWe} and
\eqref{eq:assobs2}.
Also since $
\mathcal{C}_{\hax}= \widehat{\mathfrak{C}}
$ is forward invariant, the
$\hax$-component of this solution cannot reach the boundary of this set
in finite time.\\

\noindent
\textit{{Stability}}


Let $\mathcal{N}_\ellmod$,
contained in
$\intGamma$,
 be an open neighborhood of the origin
whose points
 $(\hax,x,z)$ satisfy
\begin{eqnarray*}
& \displaystyle
V_e(x_e)\;+\;
\alpha  \left(
\alphalow ^{-1} (\sigma^d_\ellmod \;\alphaup_\ellmod(|x-\hax|))
\right)\; <\; \frac{\vv_1}{2}
\;  .
\end{eqnarray*}
Consider a
solution of the closed-loop system,
starting from an arbitrary point
$(\hax,x,z)$ in $\mathcal{N}_\ellmod$. Let $[0,T[$ be its right maximal interval of
definition when it takes its value in the open  set $\intGamma$.
To simplify the notation we add $(t)$ to denote those variables which are evaluated along this solution.

With
\eqref{eq:assobs2} and \eqref{inequalityU} we have
\begin{eqnarray*}
U(x(t),\hax(t)) &\leq &\alphaup_\ellmod(|x(0)-\hax(0)|)
\qquad \forall \;t\in [0,T[ \;,
\\
\alphalow (|x(t)-\hax(t)|) &\leq &\sigma_\ellmod^d U(x(t),\hax(t))\qquad \forall \;t\in [0,T[ \;.
\end{eqnarray*}
This implies
\be\label{inequalityTildextime}
|x(t)-\hax(t)|\leq \alphalow ^{-1} (\sigma^d_\ellmod \alphaup_\ellmod(|x(0)-\hax(0)|)) \qquad
\forall \;t\in [0,T[\;.
\ee
This inequality and \eqref{inequalityWe}, where $W_e$ is non negative, give
\begin{multline}
\label{inequalityWetime}
V_e(x(t),z(t)) \leq   V_e(x(0),z(0)) \\
+ \;
\alpha
(\alphalow ^{-1} (\sigma^d_\ellmod \;\alphaup_\ellmod(|x(0)-\hax(0)|)))
\;  < \frac{\vv_1}{2} \\
\qquad \forall \;t\in [0,T[
\  .
\end{multline}
Thus, if the  initial condition $(\hax(0), x(0),z(0))$ is in
$\mathcal{N}_\ellmod$, the solution remains inside a strict subset of
$\intGamma$.
Hence $T$ is infinite and from \eqref{inequalityTildextime} and
\eqref{inequalityWetime} we can conclude that the origin is stable.\\
\par\vspace{1em}\noindent

\textit{Attractiveness}

Consider now a solution of the closed-loop system with initial
condition $(\hax,x, z)$ in $\intChatx\times \Omega _{\vvs_1}$
which, according to (\ref{eq:LP7}), contains $\intChatx\times \mathcal{C} _{x\!z}$.
Let $[0,T[$ be
the
 right maximal interval
of definition of this solution when it takes
its values in $\intGamma$.
With (\ref{eq:assobs1}), (\ref{eq:assobs2})
and (\ref{inequalityWe}), we have, for all $t$ in $[0,T[\;$,
\begin{eqnarray}
\nonumber
U(x(t),\hax(t)) &\leq &\exp\left(-\sigma_\ellmod^d\; t\right)U(x(0),\hax(0))
\\\label{inequalityUtime}
&\leq &
\exp\left(-\sigma_\ellmod^d\; t\right)\sigma_\ellmod^{d}\, \overline{U}
\,   ,
\\[0.5em]\nonumber
V_e(x(t),z(t))& \leq &V_e(x(0),z(0))\;+\; \overline{W}\,  t
\; \leq \; \vv_1\;+\; \overline{W}\,  t
\  .
\end{eqnarray}
Since the $\hax$-component of the solution cannot reach the boundary
of $\mathcal{C}_{\hax}$ in finite time and since $V_e(x(t),z(t))$ is
smaller than $\vv_2$,
we must have
$$
T\; \geq \; \frac{\vv_2-\vv_1}{\overline{W}}
$$
and
\begin{equation}
\label{eq:LP11}
V_e\left(x\left(\dfrac{\vv_2-\vv_1}{2\overline{W}} \right),z\left(\dfrac{\vv_2-\vv_1}{2\overline{W}} \right)\right)\; \leq \;
 \frac{\vv_2+\vv_1}{2 }
\; < \; \vv_2
\  .
\end{equation}
Then, because $\ellmod $ satisfies (\ref{eq:LP16}),
(\ref{inequalityUtime}) and (\ref{eq:assobs2}) give,
for all $t$ in $\textstyle[\frac{\vvs_2-\vvs_1}{2\overline{W}} , T[\;$,
$$
\alphalow(|
x(t)-\hax(t)
|)\, \leq \,
\sigma_\ellmod^d U(x(t),\hax(t))\,\leq \,
\alphalow \left(\delta _{xw}\right)
$$
and therefore
\begin{equation}
\label{inequalityTildextime2}
|x(t)-\hax(t)|\; \leq \;   \delta _{xw}
\qquad
\forall \;t\in \textstyle[\frac{\vvs_2-\vvs_1}{2\overline{W}} , T[\displaystyle
\  .
\end{equation}
Then, with \eqref{inequalityWeconstant} and (\ref{eq:LP11}), we obtain
\begin{multline}
\label{eq:LP12}
\max\left\{
\vrule height 0.5em depth 0.5em width 0pt
V_e(x(t),z(t))\,  ,\,  \vv_1\right\}\;
\leq
\\  \;
\max\left\{
V_e\left(x\!\left(\dfrac{\vv_2-\vv_1}{2\overline{W}} \right)\!,\,z\!\left(\dfrac{\vv_2-\vv_1}{2\overline{W}} \right)\right)
\!,{{}}\vv_1
\right\} < {{}}\vv_2
\quad 
\end{multline}
for all $t$ in $[\frac{\vvs_2-\vvs_1}{2\overline{W}} , T[$.
Since $\intChatx$ is forward invariant, this establishes
that the solution cannot reach the
boundary of
$\intGamma$
 on $[0,T[$.
This implies that $T$ is infinite and that the solution remains in
$\intGamma $ for all $t$ in $\RR_{\geq 0}$. So inequalities \eqref{inequalityTildextime2} and
(\ref{eq:LP12}) and therefore inequalities \eqref{inequalityWe},
(\ref{eq:assobs2}) and \eqref{eq:assobs1}
  hold for all $t$ larger than
$\frac{\vvs_2-\vvs_1}{2\overline{W}}$.
With LaSalle invariance principle, we conclude
$$
\lim_{t\to +\infty } V_e(x(t),z(t))+U(x(t),\hax(t))\;=\; 0
\  .
$$
and thus that the solution of the closed-loop system converges to the origin
provided its initial condition $(\hax (0),x(0),z(0))$ is
in $\intChatx\times \Omega _{\vvs_1} \supset
\intChatx\times \mathcal{C} _{x\!z}\,$.
\hfill \vrule height 0.55em depth 0pt width 0.55em
} 

\subsection{Proof of Proposition \ref{prop:robust} }
\label{sec:proofRobust}

We denote
\begin{equation}
\label{eq:LP19}
\begin{array}{@{}r@{\; }c@{\; }l@{}}
\hx &=& (x,z,\hax)
\  ,
\\[0.3em]
\varphi _m(\hx) &=&\left(\begin{array}{@{}c@{}}
f(x) + g(x)\bpsi(\hax,z) 
\\
k(\hax ,h _r(x))
\\
\vartheta_\ellmod (h(x),\hax,\bpsi(\hax,z) )
\end{array}\right)
\  ,
\\[1.5em]
\varphi _p(\hx)&=&\left(\begin{array}{@{}c@{}}
\xi (x,\bpsi(\hax,z) 
\\
k(\hax ,\zeta _r(x,\bpsi(\hax,z) ))
\\
\vartheta_\ellmod (\zeta (x,\bpsi(\hax,z) ),\hax,\bpsi(\hax,z) )
\end{array}\right)\  .
\end{array}
\end{equation}

A first elementary remark is that, if 
$\hx_e=(x_e,z_e,\hax_e)$ is an equilibrium point of 
$\varphi_p$, then we have in particular
$$
0=\left.\dot z\right|_{\hx=\hx_e}=k(\hax_e,h_r(x_e))
\  .
$$
With (\ref{eq:kcond1}) this implies $h_r(x_e)$ is zero.

To prove the existence of $\hx_e$, we use
\cite[Theorem 8.2]{Hale80} which says that a forward invariant 
set which is homeomorphic to the closed unit ball of $\RR^n$ contains 
an equilibrium. The Proposition \ref{prop:outF} gives us a forward invariant set, 
which may not be homeomorphic to the closed unit ball. So our 
next task is to build another set satisfying the required properties.

The equilibrium of 
\be\label{eq:phi_m}
\dot \hx=\varphi_m(\hx)
\ee
being asymptotically attractive and interior to $\overline{\Comp}$ 
which is forward invariant, $\overline{\Comp}$ is attractive. It is also stable due 
to the continuity of 
solutions with respect to initial conditions uniformly on compact 
time subsets of the domain of definition. So it is 
asymptotically stable with the same domain of attraction $\A$ as the 
equilibrium. It follows from \cite[Theorem 3.2]{Wilson69} that there 
exist $C^\infty$ functions  
$V:\A\to \RR_{\geq 0}$ and $V_{\overline{\Comp}}:\A\to \RR_{\geq 0}$ 
which are proper on $\A$
and a class $\mathcal{K}_\infty $ function $\alpha $ 
satisfying
\begin{equation}
\label{eq:LP25}
\begin{array}{@{}c@{}}
\ba{r@{\; }c@{\; }l@{\qquad }r@{\; }c@{\; }l}
\alpha (|\hx|)& \leq & V(\hx) \;,& \quad V(0) & = & 0\;,
\\[.3em]
\displaystyle 
\alpha (d(\hx,\overline{\Comp}))& \leq & V_{\overline{\Comp}}(\hx) \;, 
&  V_{\overline{\Comp}}(\hx) & = & 0\quad \forall \, \hx \in \overline{\Comp}\;,
\end{array}\\[1.2em]
\ba{r@{\; }c@{\; }l}
\displaystyle 
\pderiv{V}{\hx}(\hx)\, \varphi_m(\hx) & \leq  & -V(\hx) \qquad \forall \, \hx \in \A \;,
\\[.5em]
\pderiv{V_{\overline{\Comp}}}{\hx}(\hx)\,\varphi_m(\hx)
& \leq  & -V_{\overline{\Comp}}(\hx) \qquad \forall \, \hx \in \A\;.
\ea
\end{array}
\end{equation}
Since 
$\overline{\Comp}$ is compact and $\mathcal{N}_{\partial \overline{\Comp}}$ is a 
neighborhood of its boundary, there exists a strictly positive real 
number $\overline{d}$ such that the set
$\{\hx\in\A\!:\,  d(\hx,\overline{\Comp}) \in (0, \overline{d}]\}$ is a subset of 
$\mathcal{N}_{\partial \overline{\Comp}}$.
Then, with the notations
$$
v_{\overline{\Comp}} \;=\;    \sup_{\hx\in\A: \,  d(\hx,\overline{\Comp})\leq \overline{d}  }V(\hx )
\quad ,\qquad 
\varpi =\frac{\alpha (\overline{d}   )}{2v_{\overline{\Comp}}}
\  ,
$$
and since $\alpha $ is of class $\mathcal{K}_\infty $, we obtain 
the implications
$$
\begin{array}[t]{@{}l@{\quad \Rightarrow \quad }l@{}}
V_{\overline{\Comp}}(\hx) \!+ \!\varpi   V(\hx) \!=\!\alpha (\overline{d}   )
&
\alpha (d(\hx,\overline{\Comp}))\!\leq\! V_{\overline{\Comp}}(\hx) \!\leq \!\alpha (\overline{d}   )
\: ,
\\[.3em] &
d(\hx,\overline{\Comp})\leq \overline{d}   
\: ,
\\[.3em] &
V(\hx )\leq v_{\overline{\Comp}}
\:   .
\end{array}
$$
With our definition of $\varpi $, this yields also
\begin{equation}
\label{eq:LP27}
\begin{array}[b]{@{}l@{\quad \Rightarrow \quad }l@{}}
\alpha (\overline{d}   )-\varpi \,   V(\hx)
= V_{\overline{\Comp}}(\hx)  
&
0\; <\; \frac{\alpha (\overline{d}   )}{2} \leq  V_{\overline{\Comp}}(\hx)
\: ,
\\[0.3em] &
0 < d(\hx,\overline{\Comp})\leq \overline{d} 
\: ,
\\[0.3em] &
\hx \in 
\mathcal{N}_{\partial \overline{\Comp}}\setminus \overline{\Comp}
\;  .  
\end{array}
\end{equation}
On the other hand, with the compact notation
$$
\Lyap (\hx) = V_{\overline{\Comp}}(\hx)+\varpi V(\hx)
\  ,
$$
we have
\begin{equation}
\label{eq:LP26}
\frac{\partial \Lyap }{\partial 
\hx}(\hx) \, \varphi _m(\hx) 
\: <\:  -\Lyap (\hx)
\qquad \forall \hx \in \A
\: .
\end{equation}
All this implies that $\Lyap$ is a Lyapunov Function for \eqref{eq:phi_m} on $\A$
in the sense of \cite[Page 324]{Wilson} and that
the sublevel set
$
\{\hx \in \A:\,  \Lyap (\hx) \leq  \alpha ( \overline{d}    )\}
$
is contained in $\mathcal{N}_{\partial \overline{\Comp}}\cup \overline{\Comp} $.
It follows from \cite[Corollary 2.3]{Wilson}\footnote{Thanks to the 
contribution of Freedman \cite{Freedman} and Perelman
\cite{Morgan-Gang}
the restriction on the dimension is not needed.}
that the level set
$
\{\hx \in \A :\,  \Lyap (\hx) = \alpha (\overline{d}   )\}
$
is homeomorphic to the unit sphere. 
But, with the fact that 
the origin is asymptotically stable and the arguments used in the 
proof of \cite[Theorem 1.2]{Wilson}, this implies that the 
sublevel set
$
\{\hx \in \A:\,  \Lyap (\hx) \leq  \alpha (\overline{d}   )\}
$ is homeomorphic to the closed unit ball.

Then, since the set
$$
C = \{\hx \in \mathcal{N}_{ \partial \overline{\Comp}} : d(\hx,\overline{\Comp}) \in [0, \overline{d}]\}
$$
is a compact subset of $\mathcal{N}_{ \partial \overline{\Comp}}\subset 
\mathcal A$, the real number
\begin{equation}
G = \sup_{\hx \in C} \left|\pderiv \Lyap \hx(\hx)\right|
\end{equation}
is well defined and strictly positive. We get, for all $\hx $ in $C$,
$$
\begin{array}{@{}r@{\; }c@{\; }l@{}}
\pderiv \Lyap  \hx(\hx) \varphi _p(\hx)&=&\displaystyle 
\pderiv \Lyap \hx(\hx) \varphi _m(\hx) + \pderiv \Lyap  \hx(\hx) [\varphi _p (\hx)-\varphi _m(\hx)]
\  ,
\\[0.7em]
&\le &\displaystyle 
-\Lyap (\hx) + G  \sup_{\hx\in C} |\varphi _p (\hx)-\varphi _m(\hx)|
\  .
\end{array}
$$
So, if $\varphi _p $ satisfies
\begin{equation}
\label{eq:LP28}
|\varphi _p(\hx) - \varphi _m(\hx)|  \le \frac{\inf_{\hx\in C} 
\Lyap (\hx)}{2G} 
\qquad \forall \hx \in 
\mathcal{N}_{\partial \overline{\Comp}}\  ,
\end{equation}
we have, for all $\hx$ in $\{\hx \in \A :\,  \Lyap (\hx) =  \alpha 
(\overline{d}   )\}$
$$
\pderiv \Lyap  \hx(\hx) \varphi _p(\hx)\leq -\frac12 \Lyap (\hx) 
\  .
$$
This implies  the compact sublevel set
$\{\hx :\,  \Lyap (\hx) \leq  \alpha (\overline{d}   )\} $
is homeomorphic to the closed unit ball and forward invariant
for the system \eqref{eq:A3}.
With \cite[Theorem 8.2]{Hale80}, we conclude that this sublevel set 
contains an equilibrium of this system.

Finally,  from points 1 and 6 of Assumption \ref{ass:observer}, we 
know that, even when
the observer in (\ref{eq:controller}) is fed with $y=\zeta 
(x,u)$ and not with $h(x)$, it admits a forward invariant compact subset 
$\widehat{\mathfrak{C}}$ of $\Obs$. So with
$$
L=\sup_{\hx \in \widehat{\mathfrak{C}}} \{ L_{\vartheta_\ellmod}(\hax) , L_k(\hax)\}
$$
with $L_{\vartheta_\ellmod}$ given by (\ref{eq:LP23}) and $L_k(\hax)$ given by \eqref{eq:kcond2}, we have,
for all $(x,z,\hax,u)$ in
$\RR^n\times\RR^r \times  \widehat{\mathfrak{C}}\times \U$,
\\[0.5em]$\displaystyle 
|\varphi _p(\hx)-\varphi_m(\hx)|
$\hfill \null \\\null \hfill $\displaystyle \leq \; 
\left|
\vrule height 0.5em depth 0.5em width 0pt
\xi(x, u) -[f(x) + g(x)u]
\right| 
\;+\; 2L
\left|
\vrule height 0.5em depth 0.5em width 0pt
\zeta (x,u)-h(x)\right| 
\  .
$\\[0.5em]
Hence (\ref{eq:LP28}) holds when (\ref{eq:LP2}) is satisfied with
$$
\delta \;=\; \frac{1}{1+2L}
\frac{
\inf_{\hx\in C} \Lyap (\hx)
}{
2\sup_{\hx \in C} \left|\pderiv \Lyap \hx(\hx)\right|
}
\  . 
$$

\hspace*{\fill}~\IEEEQED\par

\subsection{Proof of Proposition \ref{prop:robust2} }
\label{sec:proofRobust2}

We start with the following Lemma which
combines total stability and hyperbolicity and is a variation of 
\cite[Theorem 6]{PoulainPraly10}.

\begin{lemma}\label{lemma:totalstability}
Let  a $C^1$ function $\varphi _m:\RR^n\to\RR^n$ be given such that the origin is an
exponentially stable equilibrium point of:
\begin{equation}
\dot \hx = \varphi _m(\hx)
\end{equation}
with $\A$ as domain of attraction.
For any
compact sets $\underline{\Comp}$ and $\overline{\Comp} $, the latter being forward 
invariant for the above system, which satisfy
$$
\{0\}\subsetneqq \underline{\Comp}\subsetneqq\overline{\Comp} \subsetneqq\A
\  ,
$$
there exists
a strictly positive real number $\delta$ such that, for any $C^1$ function
$\varphi _p: \RR^n \to \RR^n$ which satisfies:
\begin{align}
|\varphi _p(\hx) - \varphi _m(\hx)| &\leq \delta, & \forall \hx & \in 
\overline{\Comp}, \label{eq:A1} \\
\left|\pderiv{\varphi _p}\hx(\hx) - \pderiv {\varphi 
_m}\hx(\hx)\right| &\leq \delta, & \forall \hx &\in \underline{\Comp}, \label{eq:A2}
\end{align}
there exists an exponentially stable equilibrium point  $\hx_e$ of:
\begin{equation}\label{eq:A3}
\dot\hx = \varphi _p (\hx)\  ,
\end{equation}
the basin of attraction of which contains the compact set 
$\overline{\Comp} $.
\end{lemma}

\begin{IEEEproof}
%
Let $\Pi$ be a positive definite symmetric matrix and $a$ a strictly positive real number satisfying
\[
\Pi \frac{\partial  \varphi _m }{\partial \hx} (0) +
\frac{\partial  \varphi _m }{\partial \hx} (0)^\top \Pi \; \leq -a \Pi \;,\qquad \lambda _{\min}(\Pi) = 1\;,
\]
where $\lambda _{\max}$ and $\lambda _{\min}$ respectively stand for
max and min eigenvalues.
By continuity there exists a strictly positive real number $p_0$ such
that we have, for all $\hx$ satisfying $\hx^\top \Pi \hx \le p_0\,$,
$$
\Pi \frac{\partial \varphi _m}{\partial \hx}(\hx) +
\frac{\partial \varphi _m}{\partial \hx}(\hx)^\top \Pi \le -\frac{a}{2} \Pi
$$
and
$$\
\hx^\top \Pi \varphi _m(\hx) \le -\frac a4 \hx^\top \Pi \hx.
$$

Let $\varphi _p :\RR^n\to\RR^n$ be any $C^1$ function satisfying
\begin{align}\label{eq:A4}
|\varphi _p(\hx) - \varphi _m(\hx)| &\le \frac a4 \sqrt{\frac{p_0}{12\lambda _{\max}(\Pi)}}, & \forall \hx: \hx^\top \Pi \hx &= \frac{p_0}{6}.
\end{align}
We obtain
\begin{eqnarray*}
\hx^\top \Pi \varphi _p(\hx)
&\hskip -0.5em = &\hskip -0.5em 
\hx^\top \Pi \varphi _m(\hx) + \hx^\top \Pi [\varphi _p(\hx) - \varphi _m(\hx)],
\\
&\hskip -0.5em \le &\hskip -0.5em 
\hx^\top \Pi \varphi _m(\hx) + \frac a8 \hx^\top \Pi\hx
\\[-0.3em] &\hskip -0.5em &\hskip -0.5em \quad + \frac 2a [\varphi _p(\hx) - \varphi _m(\hx)]^\top \Pi [\varphi _p(\hx) - \varphi _m(\hx)]
\end{eqnarray*}
and therefore
\begin{align}
\hx^\top \Pi \varphi _p(\hx) &\le -\frac{a}{16} \hx^\top \Pi \hx, & \forall \hx: \hx^\top \Pi \hx
&= \frac{p_0}{6}\  .
\end{align}
In this condition, it follows from \cite[Theorem 8.2]{Hale80} that, for each function $\varphi _p$
satisfying \eqref{eq:A4}, there exits a point $\hx_e$ satisfying
\begin{align}\label{eq:A5}
\varphi _p(\hx_e) &= 0, & (\hx_e)^\top \Pi \hx_e &\le \frac{p_0}{6}.
\end{align}

Assume further that $\varphi _p $ satisfies
\begin{align}\label{eq:A6}
\left|\frac{\partial \varphi _p}{\partial \hx}(\hx)
- \frac{\partial \varphi _m}{\partial \hx}(\hx)\right|
&\le \frac{a}{8 \lambda _{\max}(\Pi)}, & \forall \hx:\hx^\top \Pi \hx &\le p_0.
\end{align}
In this case, we have, for all $\hx$ satisfying $\hx^\top \Pi \hx\le 
p_0$,
\\[0.7em]$\displaystyle
\Pi \frac{\partial \varphi _p}{\partial \hx}(\hx)
+ \frac{\partial \varphi _p}{\partial \hx}(\hx)^\top \Pi =
\left[\frac{\partial \varphi _p}{\partial \hx}(\hx)
- \frac{\partial \varphi _m}{\partial \hx}(\hx)\right]^\top \Pi
$\hfill \null \\\null \hfill $\displaystyle
+\,   \Pi \frac{\partial \varphi _m}{\partial \hx}(\hx)
+ \frac{\partial \varphi _m}{\partial \hx}(\hx)^\top \Pi
+ \Pi \left[\frac{\partial \varphi _p}{\partial \hx}(\hx)
- \frac{\partial \varphi _m}{\partial \hx}\hx) \right]
$ \\$\displaystyle
\hphantom{\Pi \frac{\partial \varphi _p}{\partial \hx}(\hx)
+ \frac{\partial \varphi _p}{\partial \hx}(\hx)^\top \Pi }
\le -\frac a4 \Pi\  .
$\\[0.7em]
Note also that we have
\vspace{-1ex}
\begin{multline*}
[\hx_e + s(\hx-\hx_e)]^\top \Pi [\hx_e+s(\hx-\hx_e)] \le p_0\,, \quad
\\ \qquad
\forall (\hx,\hx_e,s) : \: s \in [0,1]\,  ,\,
 (\hx_e)^\top \Pi \hx_e \le \frac{p_0}{6}\,  ,\,
\hx^\top \Pi \hx \le  \frac{p_0}{3}.
\end{multline*}
Then, with
\begin{equation*}
\varphi _p (\hx) \!= \!\varphi _p (\hx) - \varphi _p(\hx_e)
\!=\!\! \!\int_0^1 \frac{\partial \varphi _p}{\partial \hx}(\hx_e + s(\hx - \hx_e)) ds [\hx - \hx_e]
\end{equation*}
and \eqref{eq:A5}, we get, for all $\hx$  satisfying
$\hx^\top \Pi \hx \le \frac{p_0}{3}$,
\\[0.5em]\vbox{\noindent%
$\displaystyle 
[\hx-\hx_e]^\top \Pi \varphi _p(\hx) =  
$\hfill \null \\\null \hfill $\displaystyle 
\int_0^1
\left( [\hx-\hx_e]^\top \Pi \frac{\partial \varphi _p}{\partial \hx}(\hx_e + s(\hx - \hx_e))
[\hx-\hx_e] \right)ds, 
$\\$\displaystyle 
\hphantom{[\hx-\hx_e]^\top \Pi \varphi _p(\hx)}
\le -\frac a4 [\hx - \hx_e]^\top \Pi [\hx-\hx_e]\  .
$}

Let
$$
\delta _1= \min\left\{\frac a4 \sqrt{\frac{p_0}{12\lambda _{\max}(\Pi)}}
\,  ,\,  
\frac{a}{8 \lambda _{\max}(\Pi)}
\right\}
\  ,
$$
and reduce $p_0$ if necessary to have that $\hx$ satisfying 
$(\hx_e)^\top \Pi \hx_e \le p_0$ is in
$\underline{\Comp} $. Then (\ref{eq:A1}) and (\ref{eq:A2}) with $\delta =\delta _1$ 
implies (\ref{eq:A4}) and therefore (\ref{eq:A5}). We have 
established that the system  \eqref{eq:A3}
has an exponentially stable equilibrium with basin of attraction containing
the compact set $\{\hx\in \RR^n : \hx^\top \Pi \hx \le \frac{p_0}{3}\}$.

Now, with $\overline{d}$ and $\Lyap =V_{\overline{\Comp}}+\varpi V$
as defined  in the proof of Proposition \ref{prop:robust}, we let $\underline{\vv}$ be
a strictly positive real number such that we have
\begin{equation}
\label{eq:A8} 
\hx^\top \Pi \hx \;  \le \; \frac{p_0}{3}\qquad  \forall \hx\in\A : \Lyap (\hx)  \le \underline{\vv} 
\end{equation}
Let also
$$
C = \{\hx \in \mathcal A : \underline{\vv} \le \Lyap (\hx) \; ,\;  d(\hx,\overline{\Comp}) \in [0, \overline{d}]\}
$$
It is a compact subset of $\mathcal{N}_{ \overline{\Comp}}\subset \mathcal A$.
By mimicking the same steps as in the proof of 
Proposition \ref{prop:robust}, we can obtain that, if $\varphi _p $ satisfies
\begin{equation}
\label{eq:A10}
|\varphi _p(\hx) - \varphi _m(\hx)|  \le \frac{\inf_{\hx\in C} 
\Lyap (\hx)}{2G} \  ,
\qquad  \forall \hx \in \startmod \overline{\Comp} \stopmod
\end{equation}
we have
$$
\pderiv \Lyap  \hx(\hx) \varphi _p(\hx)\leq -\frac12 \Lyap (\hx) 
\qquad  \forall \hx \in C\  .
$$
This implies  the compact set
$\{\hx\in \mathcal A : \Lyap (\hx) \le\underline{\vv} \}$
is asymptotically stable for the system \eqref{eq:A3} with
basin of attraction ${\cal B}$ containing the compact set 
$\{\hx \in\A: \Lyap (\hx)\leq \alpha (\overline{d})\}$ which contains 
$\overline\Comp$. Since, with \eqref{eq:A8}, we
have
$$
\{\hx \in \mathcal A : V(\hx) \le \underline{\vv}\} \subset \Big\{\hx \in \RR^n : \hx^\top \Pi \hx \le \dfrac{p_0}{3}\Big\}.
$$
with \eqref{eq:A4}, \eqref{eq:A6},  and \eqref{eq:A10} we have established
our result with $\delta$ given as
$$
\delta = \min\!\left\{ \frac a4 \sqrt{\frac{p_0}{12\lambda _{\max}(\Pi)}}, \frac{a}{8\lambda _{\max}(\Pi)}, \frac{\inf_{\hx \in C} V(\hx)}{2 \sup_{\hx\in C} \norm{\pderiv V \hx(\hx)}} \right\}\!\!.
$$
\end{IEEEproof}

\IEEEproofof{Proposition \ref{prop:robust2}}
In view of the above Lemma and (\ref{eq:LP22}), Proposition \ref{prop:robust2} holds if 
(\ref{OF1}) and (\ref{OF2}) imply (\ref{eq:A1}) and (\ref{eq:A2}).
At the end of the proof of Proposition \ref{prop:robust} we have seen that 
(\ref{OF1})  implies (\ref{eq:A1}).
So we are left with proving that 
(\ref{OF1}) and (\ref{OF2}) imply  (\ref{eq:A2}). 
By using again the notations (\ref{eq:LP19}) and by dropping the arguments we see that
\begin{multline*}
\left| \pderiv {\varphi_p}\hx(\hx)- \pderiv{\varphi_m}\hx(\hx)\right| \leq 
\\
\left|\Delta_{k\vartheta} (\Delta_{p}- \Delta_m) \Delta_{u} \right|
+  \left|\left(\Delta_2 + \Delta_{2\vartheta} \Delta_u \right)\right|\left|\Delta_{y} \right|,
\end{multline*}
where
\[\ba{ll}
 \Delta_{k\vartheta} = \bpmx I & 0 & 0 \\ 0 & \pderiv k {y_r}
& \pderiv {\vartheta_\ellmod} {y}
 \epmx^{\!\!\!\top} \!\!\! ,
 & 
 \Delta_{u} =  \bpmx
 I & 0 & 0
\\
0 
&   \pderiv \bpsi z 
&  \pderiv \bpsi \hax 
 \epmx\!,
\\
\Delta_{p} = \bpmx
\pderiv \xi x  &
 \pderiv \xi u 
 \\
 \pderiv \zeta x 
&
 \pderiv \zeta u 
 \epmx \!,
 &
 \Delta_{m}  = 
 \bpmx
\pderiv f x +\pderiv gx \bpsi &
g 
 \\
 \pderiv h x
&
0
 \epmx\!,
\ea\]
\[
\Delta_2 = 
\bpmx
0 & 0 & 0
\\
\dfrac{\partial^2 k}{\partial y_r^2}  \pderiv{h_r}{x}
 & 0 & \dfrac{\partial^2 k}{\partial \hax^2} 
\\
\dfrac{\partial^2 \vartheta_\ellmod}{\partial y^2}  \pderiv h x & 
0 & 
\dfrac{\partial^2 \vartheta_\ellmod}{\partial \hax^2} 
\epmx \!,
\quad
\Delta_{2\vartheta} = 
\bpmx
0 & 0 
\\
0 & 0 
\\
0& 
\dfrac{\partial^2 \vartheta_\ellmod}{\partial u^2} 
\epmx \!,
\]

$
\Delta_y = \zeta(x,\bpsi(\hax,z)) - h(x).\\
$
Recall that by construction the functions $\bpsi$, $k$ and $\vartheta$ are $C^1$. Hence, by letting
(where the arguments are dropped for compactness)
\[\ba{rcl}
\!\!
L_{k\vartheta} 
\!\! & = & \displaystyle \!\!\!\!
\sup_{(x,z,\hax) \in \underline{\Comp}}
\left\{
\left|\frac{\partial k}{\partial y_r} \right|
\,  ,\,  
\left|  \frac{\partial \vartheta_\ellmod}{\partial y}\right|
\right\},
\;
L_h  = 
\sup_{x \in (\underline{\Comp})_x}
\left\{
\left|\frac{\partial h}{\partial x}\right|
\right\},
\\
\!\!
L_u 
\!\! & = & \displaystyle \!\!\!\!
\sup_{(z,\hax) \in (\underline{\Comp})_{z,\hax}}
\left\{
\left|\frac{\partial \bpsi}{\partial z}\right|
\,  ,\,  
\left|  \frac{\partial \bpsi}{\partial \hax}\right|
\right\},
\\
\!\!
L_{2k}
\!\! & = & \displaystyle \!\!\!\!
\sup_{(x,z,\hax)\in \underline{\Comp}}
\left\{
\left|\dfrac{\partial^2 k}{\partial y_r^2}\right|
\,  ,\,  
\left| \dfrac{\partial^2 k}{\partial \hax^2}\right|
\right\},
\\
\!\!
L_{2\vartheta}
\!\! & = & \displaystyle \!\!\!\!
\sup_{(x,z,\hax) \in \underline{\Comp}}
\left\{
\left| \dfrac{\partial^2 \vartheta_\ellmod}{\partial y_r^2}\right|
\,  ,\,  
\left| \dfrac{\partial^2 \vartheta_\ellmod}{\partial u^2}\right|
\,  ,\,  
\left| \dfrac{\partial^2 \vartheta_\ellmod}{\partial \hax^2}\right|
\right\},
\ea\]
and $L_2 = \max \{L_{2k}, L_{2\vartheta}\}$,
we have, for all
$(x,z,\hax)$ in
$\underline{\Comp}$
\begin{multline*}
\left| \pderiv {\varphi_p}\hx(\hx)- \pderiv{\varphi_m}\hx(\hx)\right| \leq 
\\
4L_u L_{k\vartheta}
\left|\Delta_{p}- \Delta_m\right|
+ 2L_2(1 + L_u + L_h)   \left|\Delta_{y} \right|.
\end{multline*}
The proof can be completed  by using \eqref{OF1} and \eqref{OF2} in place 
of $\Delta_y$ and $(\Delta_p-\Delta_m)$ and by properly defining $\delta$.
\endIEEEproof

\section{Illustration of the Proposed Design via the \\ Longitudinal Model of a Plane}
\label{sec:example}

As an illustration we consider a non academic but still very simplified model of the longitudinal
dynamics of a fixed-wing vehicle flying at high speed, 
given (see \cite{PoulainThesis,PoulainPraly10}) by 
\be\label{eq:examplesystem}
\begin{array}{rcl}
\dot {v}& = &e -
g
\sin(\gamma )
\\
\dot \gamma & =&
\pounds \, 
 v
\sin(\theta -\gamma )
-\displaystyle \frac{
g \cos(\gamma )
}{v}
\\
\dot \theta & = & q
\end{array}\ee
where $v$ is the modulus of the speed, $\gamma $ is the path angle,
$\theta  $ is the pitch angle, $q$ is the pitch rate, $g$ is the
standard gravitational acceleration
and $\pounds  $ is an aerodynamic lift coefficient.
This model makes sense  for 
$v$ strictly positive only.

 The problem is to regulate $\gamma $ at $0$,
with $v$ remaining close to a prescribed cruise speed $v_0$,
using the pitch rate $q$ and the thrust
$e$ as controls, and with $\gamma $ and $\theta $ as only measurements. 
So here, by using the notation introduced in Section \ref{sec:mainresult}
$$
x=(\theta,\gamma ,v )\; ,\quad
u=(e,q )\; ,\quad
y=(\theta, \gamma )\; ,\quad
y_r=\gamma\; .
$$

\subsection{Choice of the function $k$ in the integral action}
We select
$$
k(x,h(x))\;=\;  v \sin(\gamma )\;.
$$
The motivation is that, then the integrator state $z$ has the same dynamics as the
altitude of the vehicle (not taken into account in this illustration).

\subsection{State feedback design}

To design the state feedback $\psi$ and the associated Lyapunov 
function $V_e$, we start by noting that the so called phugoid mode is
conservative \startmod(see for instance \cite[Section VII.4]{Andronov-Vitt-Khaikin}). \stopmod Precisely
we have that the following
function 
remains constant along the solutions
when $e=0$ and $\sin(\theta -\gamma )=\frac{g}{\pounds  v_0^2}$
$$
\mathcal{I}(v,\gamma )\;=\; \frac{v^3}{3 v_0^3}\;-\;
\frac{v}{v_0}\cos(\gamma )
$$
This can be checked by looking the time derivative of $\cal I$. 
Also the open sublevel set of $\mathcal{I}$ 
$$
\Stab
\;=\; 
\left\{(v,\gamma )\,  :\: \frac{v^3}{3 v_0^3}\;-\;
\frac{v}{v_0}\cos(\gamma ) < 0\right\}
$$
is the largest sublevel set not containing a point of the type 
$(0,\gamma )$. Namely it is the largest sublevel set of $\mathcal{I}$ where the model (\ref{eq:examplesystem}) is well 
defined. Moreover in this set the $\gamma $-component of any point is 
in $\left(-\frac{\pi }{2},\frac{\pi }{2}\right)$.
Also $\mathcal{I}$ is positive definite in
$v-v_0$ and $\gamma $ on $\Stab $. We conclude that 
$\mathcal{I}+\frac{2}{3}$, restricted to $\Stab $ is a candidate for playing 
the role of a Lyapunov function.
Also forwarding with the functions $V$ and $H$ known is possible since
when $e=0$, the function
$$
z\;-\; H(v)\;=\;  z+
\frac{v^2}{2g}
$$
remains constant along the solutions of the following
$(z,v)$-subsystem
\begin{eqnarray*}
\dot z&=&v\,\sin(\gamma )
\\
\dot v&=&e-g\,  \sin(\gamma )\  .
\end{eqnarray*}

Finally we can
complete the design of a state feedback by applying
backstepping from the fact
that $\theta $ given as
$$
\theta \;=\; \gamma \;+\; \arcsin\left(\frac{g}{\pounds  v_0^2}\right)
$$
is stabilizing for the $(z,v,\gamma )$-subsystem.

All this leads to the following (weak)\footnote{%
Its derivative along the solutions may be only non positive.}
Control Lyapunov function 
\begin{multline} \label{ex1}
V(z,v,\gamma )
=
\frac{ v^3}{3v_0^3}\:+\: \frac{2}{3}\:-\:
\frac{v}{v_0}\,\cos(\gamma )
\:+\: \frac{k _1}{4}\left(
\frac{2gz + v^2-v_0^2}{v_0^2}
\right)^2 \\
+ \frac{k _2}{2}
\left[\theta -\gamma -\arcsin\left(\frac{g}{\pounds  v_0^2}\right)\right]^2
,
\end{multline}
where
 the dimensionless numbers $k _1$ and 
$k _2$ are arbitrary but strictly positive, 
and the following feedback law
\\[1em]$\displaystyle 
    e \;=\;  -\texttt{sat}_{k_e}\left(k _3\left[
\frac{v^2}{v_0^2}-\cos(\gamma)
+k _1
\frac{2gz + v^2-v_0^2}{v_0^2}
\frac{v}{v_0}\right]
\right), $\hfill \null 
\\[1em]$\displaystyle\displaystyle 
q \;=\;  -\left(
\frac{\pounds  v_0^2\sin(\theta -\gamma ) -g
}{\theta -\gamma -\arcsin\left(\frac{g}{\pounds  v_0^2}\right)}
\frac{v^2}{k _2v_0^3}\sin(\gamma ) + \dfrac{g\cos(\gamma )}{v}
\right.
$\hfill \null \\[0em]\null \hfill $\displaystyle 
\left.
\vrule height 1.5em depth 1.5em width 0pt
- \pounds  v
\sin(\theta -\gamma )
+k _4 \!\left[\theta -\gamma -\arcsin \!\left(\frac{g}{\pounds  v_0^2}\right)\right]
\!\right)\!,
$\\[.5em]
where $k _3$ and $k _4$ are dimensionless arbitrary strictly positive real numbers
and $k_e$ and $k_q$ are arbitrary strictly positive saturation levels.
With LaSalle
invariance principle 
 it is possible to prove that $(v,\gamma,\theta) = \left(v_0,0,\arcsin\left( \frac{g}{\pounds v_0^2}\right)\right)$ is the only
asymptotically stable equilibrium point of the system \eqref{eq:examplesystem}. 
In this simple illustration we have chosen the simpler Lyapunov Function \eqref{ex1}, but it does not give enough degrees of freedom to improve 
performance and increase the domain of attraction. 
More appropriate designs are possible 
by choosing different Lyapunov functions (see \cite{PoulainThesis}). 
Finally, according to Section \ref{sec:designProp}, for its use in the output feedback, the state feedback law $q$ above has to be modified 
by adding a saturation (see in particular the function $\bpsi$ in \eqref{eq:bpsi}).


\subsection{Design of the high-gain observer}
To obtain an observer we check that the conditions of
Assumptions \ref{ass:existHGO} are satisfied. 
Let $\gamma _{dot}$ be defined as the following function
$$
\gamma _{dot}(\theta ,\gamma ,v)= \pounds 
v\sin(\theta -\gamma )-
\frac{g \cos(\gamma)}{v}
$$
Then let
$$
\Phi((\theta ,\gamma ,v))=\Phi(x)= (\phi _1,\phi _2,\phi _3)=(\theta,
\gamma, \gamma _{dot}(\theta ,\gamma ,v))
\  .
$$
It is defined on the set
$$
\Obs  =  \left(-\frac \pi 2;\frac \pi 2\right)\times \left(-\frac
\pi 2;\frac \pi 2\right)\times (0;+\infty)\;,
$$
and $(\theta ,\gamma ,v)$ can be recovered from its values
$(\phi _1,\phi _2,\phi _3)$ in the following 
subset\footnote{%
We use  $|\phi _1-\phi _2|$ to upper bound $\cos\phi _2\sin(\phi _1-\phi _2)$.} 
of $\Phi(\Obs )$
\begin{multline*}
\Xi \;=\;  \Big\{\phi \in\RR^3 : \phi _1\in\left(-\frac\pi2;\frac\pi2\right),  \; \phi _2\in\left(-\frac\pi2;\frac\pi2\right),
\\
\phi _3 < -2\sqrt{g
\pounds 
|\phi _1-\phi _2|} \quad \mbox{if} \quad (\phi _1-\phi _2)\leq0\Big\}\;.
\end{multline*}
Note also that $\partial \Phi / \partial x$
is always non-singular on the set $\Obs $ because ${\partial \gamma _{dot}}/{\partial  v}$ cannot be equal to 0 when
$\phi \in\Xi$.
Hence the function $\Phi$ is
a diffeomorphism satisfying
 Assumption O1.

Then, with $C$ defined as
$$
C = \renewcommand{\arraystretch}{1} \begin{pmatrix} 1 & 0 & 0 \\ 0 & 1 & 0 \end{pmatrix},
$$
Assumption O2 
also holds. 

Now let $A,B,L_\ell,M_\ell$ and $N_\ell$ be 
defined as
$$
A =  \renewcommand{\arraystretch}{1.2} 
\begin{pmatrix} 0 & 0 & 0 \\ 0 & 0 & 1 \\ 0 & 0 & 0 \end{pmatrix} 
\quad
\ba{rcl}
L_\ell & = & \text{diag}(1,1,\ell)\ ,
\\
M_\ell & = & \text{diag}(\ell,\ell,\ell) \ ,
\\
N_\ell & = & \text{diag}(1,1)\;,
\ea
$$
\begin{multline*}
B(\Phi(x),u) = {\rm col} \left(u_1, \; 0, \;  
\frac{\partial \gamma _{dot}}{\partial \theta }u_1 + 
\frac{\partial  \gamma _{dot}}{\partial v }u_2
\right.
\\
\left.
+\frac{\partial  \gamma _{dot}}{\partial \gamma}
\gamma _{dot} - 
\frac{\partial  \gamma _{dot}}{\partial v }
g\sin (\gamma) \right)\;.
\end{multline*}
Also, given
 any strictly positive number $\nu$, let $P$ be a symmetric positive definite matrix defined as
$$ \renewcommand{\arraystretch}{1.1}
P = \begin{pmatrix} * & * & * \\ * & * & p_{23} \\ * & p_{23} & p_{33} \end{pmatrix}
$$
where $2p_{23}\leq -\nu p_{33}$. Then
there exists a real number $\rho$ such that we have
$$
PA+A^\top P- \rho \; C^\top C\; \leq \; -\nu P
\  .
$$
This implies the existence of a real number
$\underline {\nu_k}$ such that, for any $\nu_k\geq \underline
{\nu_k}$,
with
$$
K = \nu_kP^{-1}C^\top
$$
assumptions  O3 to O7
are satisfied.

\subsection{Design of the correction term}
Following
  Section \ref{sec:observer}, the function $h_2(x)$ can be defined as
$$
h_2 (x) =
h_{2}^1(x) + h_{2}^2(x) + h_{2}^3(x) + h_{2}^4(x)
$$
with
\[\ba{l}
h_{2}^1(x) = \max \left\{\frac{4\theta^2}{\pi^2}-\varepsilon_1 ; 0\right\}^{\!2}\!\!, 
\;\;
h_{2}^2(x) = \max \left\{\frac{4\gamma^2}{\pi^2}-\varepsilon_2 ; 0\right\}^{\!2}\!\!,
\\
h_{2}^3(x) = \max \Big\{
\varepsilon_3\,(\theta-\gamma)-\gamma _{dot}-\varepsilon_4
; 0\Big\}^{\!2}\!\!,
\\
h_{2}^4(x) = \max \left\{\frac{\gamma _{dot}}{\gamma _{dot\,  \max{}}} - \varepsilon_5 ; 0\right\}^{\!2}\!\!,
\ea\]
where $\varepsilon_1, \varepsilon_2,\varepsilon_3 ,\varepsilon_4,\varepsilon_5$ and $\gamma _{dot\,  \max{}}$ are constants to be properly chosen.
The functions $h_{2,1}$ and $h_{2,2}$ take care respectively of $\theta$ and $\gamma$ to stay in the set $\Xi$ as showed in Figure \ref{fig:h212}, whereas
functions $h_{2,3}$ and $h_{2,4}$ take care of $f(\theta,\gamma, v)$ as in Figure \ref{fig:h234}.

The correction term $E$ is defined as in Lemma \ref{prop:observer}.
Finally the functions $U_\ellmod$ and $\sigma _\ellmod$ can be defined 
as in the proof of Lemma \ref{lem:observerR}.

\setlength{\unitlength}{5cm}
\begin{center}
\begin{picture}(1.6,1)(-.25,0)

\put(0,0.5){\vector(1,0){1}}
\put(0.9,.4){$\theta$}
\put(.5,0){\vector(0,1){1}}
\put(.4,0.95){$\gamma$}

\put(.2,0.15){\line(0,1){0.7}}
\put(.8,0.15){\line(0,1){0.7}}

\put(.15,0.2){\line(1,0){0.7}}
\put(.15,0.8){\line(1,0){0.7}}

\put(.1,.92){\vector(3,-2){0.09}}
\put(-.25,.91){$h_{2}^1(x)=0$}
\put(.89,.92){\vector(-3,-2){0.09}}
\put(.91,.91){$h_{2}^1(x)=0$}
\put(.98,.8){\vector(-1,0){0.09}}
\put(1.,.78){$h_{2}^2(x)=0$}
\put(.98,.2){\vector(-1,0){0.09}}
\put(1,.18){$h_{2}^2(x)=0$}

\put(.2,0.5){\circle{0.01}}
\put(.8,0.5){\circle{0.01}}
\put(.5,0.2){\circle{0.01}}
\put(.5,0.8){\circle{0.01}}

\put(.09,.53){$-\frac\pi2$}
\put(.82,.53){$\frac\pi2$}
\put(.52,.12){$-\frac\pi2$}
\put(.52,.84){$\frac\pi2$}
\put(.63,.33){$\Xi$}


\put(.3,0.22){\line(-1,1){0.075}}
\put(.395,0.22){\line(-1,1){0.075}}
\put(.49,0.22){\line(-1,1){0.075}}
\put(.585,0.22){\line(-1,1){0.075}}
\put(.68,0.22){\line(-1,1){0.075}}
\put(.775,0.22){\line(-1,1){0.075}}

\put(.3,0.315){\line(-1,1){0.075}}
\put(.395,0.315){\line(-1,1){0.075}}
\put(.49,0.315){\line(-1,1){0.075}}
\put(.585,0.315){\line(-1,1){0.075}}
\put(.775,0.315){\line(-1,1){0.075}}

\put(.3,0.41){\line(-1,1){0.075}}
\put(.395,0.41){\line(-1,1){0.075}}
\put(.49,0.41){\line(-1,1){0.075}}
\put(.585,0.41){\line(-1,1){0.075}}
\put(.68,0.41){\line(-1,1){0.075}}
\put(.775,0.41){\line(-1,1){0.075}}

\put(.3,0.505){\line(-1,1){0.075}}
\put(.395,0.505){\line(-1,1){0.075}}
\put(.49,0.505){\line(-1,1){0.075}}
\put(.585,0.505){\line(-1,1){0.075}}
\put(.68,0.505){\line(-1,1){0.075}}
\put(.775,0.505){\line(-1,1){0.075}}

\put(.3,0.6){\line(-1,1){0.075}}
\put(.395,0.6){\line(-1,1){0.075}}
\put(.49,0.6){\line(-1,1){0.075}}
\put(.585,0.6){\line(-1,1){0.075}}
\put(.68,0.6){\line(-1,1){0.075}}
\put(.775,0.6){\line(-1,1){0.075}}

\put(.3,0.695){\line(-1,1){0.075}}
\put(.395,0.695){\line(-1,1){0.075}}
\put(.49,0.695){\line(-1,1){0.075}}
\put(.585,0.695){\line(-1,1){0.075}}
\put(.68,0.695){\line(-1,1){0.075}}
\put(.775,0.695){\line(-1,1){0.075}}
\end{picture}
\end{center} 
\vspace{-3ex}

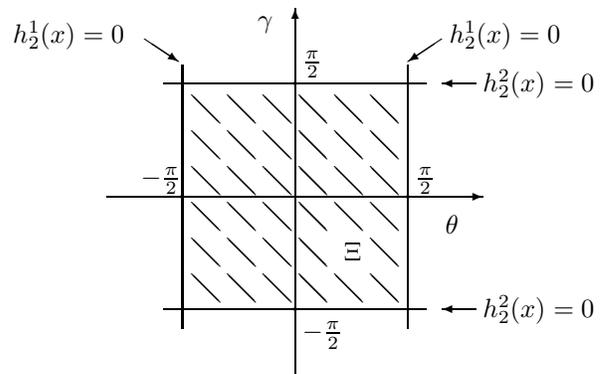
\captionof{figure}{Design of the functions $h_{2}^1\,,\; h_{2}^2\,$.}\label{fig:h212}

\setlength{\unitlength}{7cm}
\begin{center}
\begin{picture}(1.14,1)(0,0)

\put(0,0.5){\vector(1,0){1}}
\put(0.9,.45){$\theta-\gamma$}
\put(.5,0){\vector(0,1){1}}
\put(.42,0.95){$\gamma _{dot}$}

\put(.25,0.5){\circle{0.01}}
\put(.75,0.5){\circle{0.01}}
\put(.17,.52){$-\pi$}
\put(.77,.52){$\pi$}

\qbezier(0.5, 0.5)(.5, .4)(0, 0.3)
\put(.25,0.05){\line(0,1){0.31}}
\put(.75,0.05){\line(0,1){0.9}}

\put(0.2,.15){\line(1,0){.6}}
\put(.87,.2){\vector(-1,-1){0.051}}
\put(.89,.2){$h_{2}^4(x)=0$}

\put(0.55,0.5){\line(5,4){.25}}
\put(0.55,0.5){\line(-5,-4){.35}}
\put(.87,.76){\vector(-1,-1){0.051}}
\put(.89,.76){$h_{2}^3(x)=0$}

\put(.6,.29){$\Xi$}


\put(.32,0.04){\line(-1,1){0.051}}
\put(.40,0.04){\line(-1,1){0.051}}
\put(.48,0.04){\line(-1,1){0.051}}
\put(.56,0.04){\line(-1,1){0.051}}
\put(.64,0.04){\line(-1,1){0.051}}
\put(.72,0.04){\line(-1,1){0.051}}

\put(.32,0.12){\line(-1,1){0.051}}
\put(.40,0.12){\line(-1,1){0.051}}
\put(.48,0.12){\line(-1,1){0.051}}
\put(.56,0.12){\line(-1,1){0.051}}
\put(.64,0.12){\line(-1,1){0.051}}
\put(.72,0.12){\line(-1,1){0.051}}

\put(.32,0.2){\line(-1,1){0.051}}
\put(.40,0.2){\line(-1,1){0.051}}
\put(.48,0.2){\line(-1,1){0.051}}
\put(.56,0.2){\line(-1,1){0.051}}
\put(.64,0.2){\line(-1,1){0.051}}
\put(.72,0.2){\line(-1,1){0.051}}

\put(.32,0.28){\line(-1,1){0.051}}
\put(.40,0.28){\line(-1,1){0.051}}
\put(.48,0.28){\line(-1,1){0.051}}
\put(.56,0.28){\line(-1,1){0.051}}
\put(.72,0.28){\line(-1,1){0.051}}

\put(.48,0.36){\line(-1,1){0.051}}
\put(.56,0.36){\line(-1,1){0.051}}
\put(.64,0.36){\line(-1,1){0.051}}
\put(.72,0.36){\line(-1,1){0.051}}

\put(.56,0.44){\line(-1,1){0.051}}
\put(.64,0.44){\line(-1,1){0.051}}
\put(.72,0.44){\line(-1,1){0.051}}

\put(.56,0.52){\line(-1,1){0.051}}
\put(.64,0.52){\line(-1,1){0.051}}
\put(.72,0.52){\line(-1,1){0.051}}

\put(.56,0.60){\line(-1,1){0.051}}
\put(.64,0.60){\line(-1,1){0.051}}
\put(.72,0.60){\line(-1,1){0.051}}

\put(.56,0.68){\line(-1,1){0.051}}
\put(.64,0.68){\line(-1,1){0.051}}
\put(.72,0.68){\line(-1,1){0.051}}

\put(.56,0.74){\line(-1,1){0.051}}
\put(.64,0.74){\line(-1,1){0.051}}
\put(.72,0.74){\line(-1,1){0.051}}

\put(.56,0.82){\line(-1,1){0.051}}
\put(.64,0.82){\line(-1,1){0.051}}
\put(.72,0.82){\line(-1,1){0.051}}

\put(.56,0.90){\line(-1,1){0.051}}
\put(.64,0.90){\line(-1,1){0.051}}
\put(.72,0.90){\line(-1,1){0.051}}

\end{picture}
\end{center}
\vspace{-3ex}

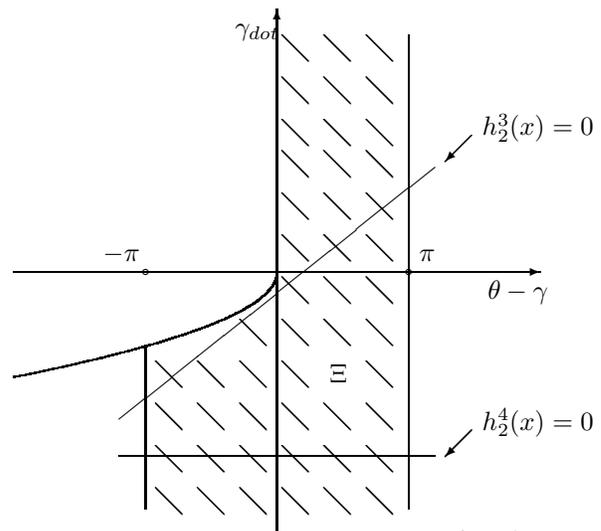
\captionof{figure}{Design of the functions $h_{2}^3\,,\; h_{2}^4\,$.}\label{fig:h234}

\section{Conclusions}
Robust asymptotic output regulation by output feedback 
has been investigated.
Our design technique follows the very usual approach of
stabilizing the origin of the model augmented with 
integrators of the output errors. 
To do so we assume we have already a stabilizing state feedback for 
the model but not asking for any specific structure 
nor for normal form nor for minimum phase. 
For the augmented model we redesign the state feedback 
by applying forwarding.
The output
feedback is obtained by introducing a high-gain observer expressed in the original
coordinates.
The output regulation is shown to be
robust to any small enough (in a $C^1$ sense) unstructured discrepancy between model
and process in open loop.

In establishing our main propositions we obtained 
new results, which may have their own interest. They concern
high-gain observers
for multi-output systems (Lemma \ref{lem:observerR}) and 
persistence of equilibria
under small perturbations (Proposition \ref{prop:robust}).

The design we propose
is illustrated by the regulation of the
flight path angle for
a simplified
 longitudinal model of a plane.

\end{document}